\UseRawInputEncoding
\documentclass[aip,reprint]{revtex4-1}

\usepackage{graphicx}
\usepackage{epsfig}
\usepackage[caption=false]{subfig}
\usepackage{amssymb}
\usepackage{amsmath}
\begin{document}
\title{Charge sensitivity of a cavity-embedded Cooper pair transistor \\limited by single-photon shot noise\\
%Charge sensitivity of a cavity-embedded Cooper pair transistor in the quantum regime bounded by photon shot-noise%
}
\author{S. Kanhirathingal}
\email{Sisira.Kanhirathingal.GR@dartmouth.edu}
\affiliation{Department of Physics and Astronomy, Dartmouth College, Hanover, New Hampshire 03755, USA} 
\author{B. L. Brock}
\affiliation{Department of Physics and Astronomy, Dartmouth College, Hanover, New Hampshire 03755, USA} 
\author{A. J. Rimberg}
\affiliation{Department of Physics and Astronomy, Dartmouth College, Hanover, New Hampshire 03755, USA} 
\author{M. P. Blencowe}
\email{Miles.P.Blencowe@dartmouth.edu}
\affiliation{Department of Physics and Astronomy, Dartmouth College, Hanover, New Hampshire 03755, USA}
\date{\today}

\begin{abstract}
Using an operator scattering approach, we analyze the quantum dynamics of an ultrasensitive electrometer -- a Cooper pair transistor embedded in a quarter-wave microwave cavity (cCPT).  
%This approach gives a first principles description of the CPT-renormalized cavity resonant frequency dependence on the applied gate voltage and magnetic flux biases. The approach furthermore validates a simpler model of the cCPT based on distributed network theory.
While the cCPT is inherently a tunable, strongly nonlinear system affording a diverse range  of functionalities, we restrict our present analysis to a necessary first investigation of its linear charge sensing capabilities, limiting to low pump powers corresponding to an average cavity photon number $\lesssim 1$. Assuming realizable cCPT parameters %similar to those obtained from a characterization of an actual experimental device%
(B. L. Brock {\it et al}., Phys. Rev. Applied {\bf 15}, 044009), and not including  noise from the subsequent amplifier chain,
we predict the fundamental, photon shot noise-limited charge sensitivity to be $0.12\, \mu{\mathrm{e}}/\sqrt{\mathrm{Hz}}$ when the pumped cavity has an average of one photon and the cCPT is operated close to charge degeneracy. This is to be compared with a first reported charge sensitivity value $14\, \mu{\mathrm{e}}/\sqrt{\mathrm{Hz}}$ in the  single-photon regime (B. L. Brock {\it et al}., arXiv:2102.05362). 
\end{abstract}
\maketitle
\section{Introduction}
 Rapid detection of electrical charge on the scale of an individual electron has long been an important experimental technique in such areas as readout of qubits,\cite{Lehnert:2003,Pla:2012} detection of individual tunneling events,\cite{Naaman:2006} and motion sensing of nanomechanical resonators.\cite{LaHaye:2004} The most common means of performing such measurements consists of detecting changes in the current flowing through a mesoscopic charge detector, such as a single electron transistor or quantum point contact, due to changes in the detector conductance.\cite{Schoelkopf:1998,Lehnert:2003,LaHaye:2004,Pla:2012}  Numerous studies have investigated the limits on the charge sensitivity, which is determined by electronic shot noise in the detector current, and where the backaction on the measured system often exceeds the minimum required by quantum mechanics.\cite{devoret-amplifying-2000,Korotkov:2003,Clerk:2004,clerk-introduction-2010}
 
 An alternative and potentially superior mode of charge detection instead relies on detecting changes in the capacitive or inductive reactance of a superconducting device such as a Cooper pair box or Cooper pair transistor that is biased on its supercurrent branch.\cite{sillanpaa-direct-2005,Persson:2010,tosi-design-2019,sillanpaa2004inductive}  By embedding such a device in a resonant circuit and measuring changes in the phase of a reflected microwave probe signal, it is possible to dispersively detect single electronic charges with a sensitivity that is limited by photon shot noise in the probe signal and with backaction  on the measured charge that may approach the minimum allowed by quantum mechanics.\cite{Zorin:2001}
 
In this paper we theoretically investigate the cavity-embedded Cooper pair transistor (cCPT),\cite{rimberg-cavity-cooper-2014,brock-ccpt-2020,brock2021fast} which functions as the first amplifier stage of a dispersive electrometer due to its charge-dependent superconducting reactance.  We show that this device is in principle capable of achieving charge sensitivities that improve upon the best predicted values for single Cooper pair transistors (SCPTs).\citep{sillanpaa-charge-2005} This is despite using many orders of magnitude less power than is typical for previous electrometer devices, in particular  corresponding to an average cavity photon number occupation $\lesssim 1$ for our cCPT device.\cite{brock-ccpt-2020,brock2021fast} While the ideal cCPT can operate as a quantum, photon shot noise-limited electrometer, the actual device in a realizable measurement setup is prone to charge fluctuations and other reducible noise sources, to date limiting its linear charge sensitivity to values two orders of magnitude worse\citep{brock2021fast} than the theoretically attainable minimum predicted in this paper. Nevertheless, the charge fluctuations can potentially be suppressed using feedback techniques that filter out the low frequency noise tampering resonance (up to a bandwidth of $\sim 10$ kHz), bringing the linear charge sensitivity of the cCPT closer to the photon shot noise-limit (not including the  noise of the subsequent amplifier chain).\citep{poundlocking_print}
 %In Sec. \ref{circuit} we begin with a simple circuit analysis of the cCPT, based on the standard modeling of distributed networks as equivalent lumped element circuits.
 
We shall   utilize a first principles, operator scattering  approach for investigating the cCPT quantum dynamics 
 that overcomes the limitations of the analyses presented 
 in Refs. \onlinecite{rimberg-cavity-cooper-2014,brock-ccpt-2020}. In particular, the present approach crucially provides the quantitative conditions under which the approximate eigenfunction expansion analysis of Ref. \onlinecite{rimberg-cavity-cooper-2014} and lumped element circuit analysis of Ref. \onlinecite{brock-ccpt-2020} are valid. Furthermore, the scattering method provides a systematic way to derive the expressions for the various parameters of the effective cavity Hamiltonian. Relevant parameters include not only those for the effective linear cavity dynamics (e.g., renormalized resonant frequency), but also the explicit forms of the higher order nonlinear cavity terms and coupling terms between the cavity and other systems such as a nanomechanical resonator.\cite{rimberg-cavity-cooper-2014} Most importantly, the scattering approach yields the versatile quantum Langevin equation for describing the effective cavity quantum dynamics, with explicit expressions for the damping and the associated quantum noise terms that are necessary for establishing the photon shot noise-limited charge sensitivity.

 As a result of its single-photon-level charge sensitivity, the cCPT is capable of mediating the standard optomechanical interaction in the ultrastrong coupling regime [see Eq. (\ref{optomech_ham_eqn}) in Sec. \ref{conclusion}]. The experimental realization of single photon optomechanical dynamics in this tripartite system (comprising the cavity, CPT, and mechanical resonator) will depend on the optimized non-linear charge sensitivity of the cCPT. While the present work does not take into account such a measured quantum dynamical system and the effects of backaction, it instead considers a deterministic sinusoidal charge modulated signal in the photon shot-noise limit as a necessary step towards such investigations.    % they are insufficient to fully understand the behaviour of the device as a charge detector in the low-photon limit.
 
 The layout of our paper is as follows. In Sec. \ref{bare}, we give a pedagogical introduction to the quantum scattering method (which is based on the superconducting circuit analysis methods introduced in Refs.\cite{yurke-quantum-1984,vool-introduction-2017}) by applying it to a bare cavity system (i.e., without the CPT). We next  derive the CPT-induced, effective cavity Hamiltonian in Sec. \ref{cCPT}. In Sec. \ref{chargesens}, we obtain the photon shot-noise limited charge sensitivity of the device when operated as a linear electrometer.
 Section \ref{conclusion} gives concluding remarks, in particular how we might define a standard quantum limit of charge sensitivity that accounts for measurement backaction (relevant for coupling the CPT to a mechanical resonator, for example), as well as how device imperfections and amplifier noise prevent the cCPT from reaching this limit. The appendices give further details of our analysis, including the approximate, lumped element circuit model description for completeness.

\section{Bare cavity-transmission line dynamics}
\label{bare}
\begin{figure}[thb]
\centering
\subfloat{\label{barec_scheme}\includegraphics[width=0.45\textwidth]{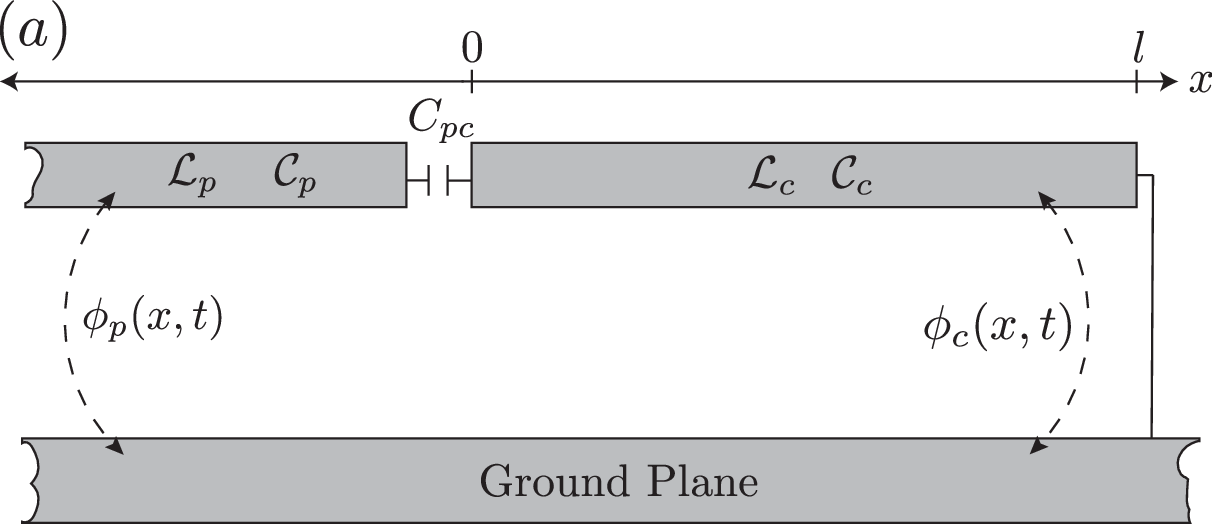}}\\
\subfloat{\label{ccpt_scheme} \includegraphics[width=0.45\textwidth] {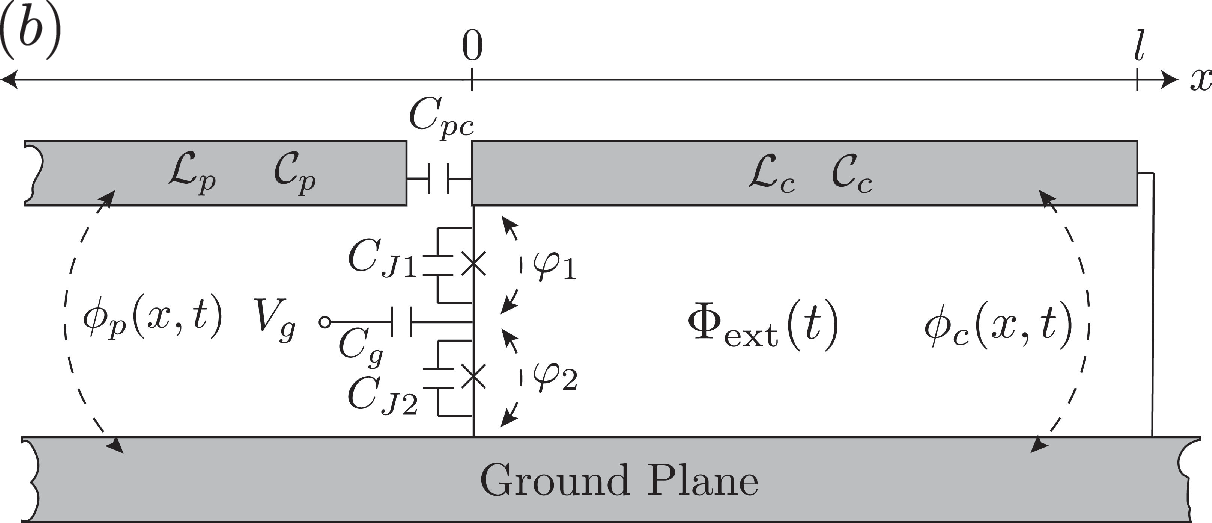}}
\caption{(a) Circuit schematic of a bare quarter-wave ($\lambda/4$) cavity coupled to pump/probe transmission line via a coupling capacitor $C_{pc}$. (b) cCPT circuit schematic.}\label{scheme}
\end{figure}
The cavity-embedded Cooper pair transistor (cCPT) consists of a shorted quarter-wave ($\lambda/4$) resonator in a co-planar wave guide geometry, and a Cooper pair transistor (CPT) at the voltage anti-node (Fig \ref{ccpt_scheme}).
%As mentioned in the previous section,
Since the CPT is designed to weakly interact with the cavity, its influence on the latter can be treated perturbatively within the operator scattering approach described later below. We shall first consider a bare cavity coupled to the pump/probe transmission line via the capacitance $C_{pc}$ in the absence of the CPT (Fig. \ref{barec_scheme}). The dynamics of this simpler `warm-up' model is well-established using the input-output formalism, with the damping rates due to internal losses and coupling to the transmission line usually considered as phenomenological parameters.\citep{gardiner-input-1985} In the following, we shall instead apply the operator scattering approach,\cite{yurke-quantum-1984,vool-introduction-2017} where we systematically recover the discrete mode cavity operators that define the cavity Hamiltonian, together with the cavity mode renormalized frequencies and external damping rates due to the coupling to the transmission line. This approach validates the lumped element circuit analysis given in Appendix \ref{circuit}. Damping due to internal losses will be neglected (i.e.,  $\kappa_{\text{int}} = 0$), to be added  phenomenologically later in Sec. \ref{chargesens}. 

It is worthwhile mentioning that the sources of the internal losses relevant to the scope of this paper originate from the interactions of the cavity with its local environment.\cite{de-graaf-suppression-2018,wang2009improving,barends2010minimal} In practice, there also exist sources of dephasing via microscopic two level system (TLS) degrees of freedom located in the vicinity of the CPT, for example within the underlying substrate and Josephson tunnel junction oxide layers. These defects couple  via their electric and magnetic dipole moments to the cCPT system charge and flux coordinates.\citep{paladino-mathbsf1mathbsfitf-2014, grabovskij2012strain, astafiev2006temperature} Such interactions are manifested as cavity resonance frequency fluctuations in the experiments;\citep{brock-ccpt-2020}  it is crucial to take these fluctuations into account when characterizing the experimental device performance since they can be erroneously equated with  additional damping.\cite{brock-frequency-2020}

To outline, we begin by writing down the  cavity and transmission line wave equations, along with the capacitive coupling and shorted-end boundary conditions using Kirchhoff's laws. The general solutions to the corresponding quantum Heisenberg wave equations that are coupled via these boundary conditions are obtained using the operator scattering approach. Under the condition of weak coupling, the standard form input-output quantum Langevin equation for the cavity mode operator is recovered by approximation, together with explicit expressions for  the resonant frequency and damping rate in terms of the circuit parameters.

\subsection{Scattering analysis}
\label{fullanalysis}
Referring to Fig. \ref{barec_scheme}, the wave equations for the cavity phase field $\phi_c(x,t)$ and the transmission line probe phase field $\phi_p(x,t)$ are
\begin{equation}
\frac{\partial^2\phi_i}{\partial t^2}=\left({\mathcal{L}_i}{\mathcal{C}_i}\right)^{-1} \frac{\partial^2\phi_i}{\partial x^2}, \, 
\begin{cases}
i = c, &\text{if}\ 0 < x < l \\
i = p, &\text{if}\ x < 0,
\end{cases}
\label{phieq}
\end{equation}
where the phase field is defined in terms of the magnetic flux field $\Phi(x,t)$ through $\phi_i \equiv 2\pi \Phi/\Phi_0$ with $\Phi_0=h/(2e)$ the flux quantum, ${\mathcal{L}}_i$, ${\mathcal{C}}_i$ denote respectively the inductance and capacitance per unit length of the cavity ($i=c$) and transmission line ($i=p$), and $l$ is the cavity center conductor length.
Current conservation at $x=0$ and the  boundary condition at $x=l$ give respectively:
\begin{equation}
\frac{1}{{\mathcal{L}_p}}\left.\frac{\partial\phi_p}{\partial x}\right|_{x=0^-}= \frac{1}{{\mathcal{L}_c}}\left.\frac{\partial\phi_c}{\partial x}\right|_{x=0^+}=C_{pc}\left.\left(\ddot{\phi}_c-\ddot{\phi}_p\right)\right|_{x=0} ,
\label{cpjunctioneq}
\end{equation}
\begin{equation}
\phi_c(l,t)=0.
\label{bclTeq}
\end{equation}

Working with the Heisenberg equations resulting from formally replacing the coordinates with  their  associated  quantum  operators $\hat{\phi}_c(x,t)$ and $\hat{\phi}_p(x,t)$, the general solution for the wave equation  (\ref{phieq}) can be written in terms of photon creation/annihilation  operators as follows:
\begin{equation}
\begin{split}
\phi_i (x,t)= & \frac{2\pi}{\Phi_0}\int_0^{\infty} d\omega  \sqrt{\frac{\hbar Z_i}{\pi \omega}}\frac{1}{2}\left[e^{-i\omega(t-t_0 -x/v_i)} a^{\to}_i (\omega,t_0) \right.\\
& \left. +e^{-i\omega(t-t_0 +x/v_i)} a^{\gets}_i(\omega,t_0)\right] +{\mathrm{h.c.}},
\label{phasepsolneq}
\end{split}
\end{equation}
where `h.c.' denotes the Hermitian conjugate and we have dropped the hats on the operators for notational convenience. Note that there should properly be a regularizing, upper frequency cut-off in Eq. (\ref{phasepsolneq}).  However, the actual measured quantities involve finite frequency bandwidths about the pump frequency that are well below (and independent of) the cut-off. The superscripts `$\to$' (`$\gets$') correspond to right (left) propagating modes, with  the photon creation/annihilation operators satisfying the standard commutation relation
\begin{equation}
[a^m_i(\omega,t_0),(a^n_i(\omega',t_0))^{\dag}]=\delta_{mn}\delta(\omega-\omega'),
\label{apcreq}
\end{equation}
where $m, n \in \{`\to\text{'}, `\gets \text{'} \}$. The cavity and transmission line impedances are given by $Z_i=\sqrt{{\mathcal{L}}_i/{\mathcal{C}}_i}$ [note $Z_i=Z_0$ in Eq. (\ref{input_impedance_eqn_quarterwave})],  and $v_i=({\mathcal{L}}_i{\mathcal{C}}_i)^{-1/2}$ is the microwave phase field propagation velocity.

In essence, the operator scattering approach involves substituting the wave equation solutions (\ref{phasepsolneq}) into the boundary conditions (\ref{cpjunctioneq}) and 
(\ref{bclTeq}) in order to express the left propagating  (i.e., ``reflected" or ``scattered") probe operator $a_p^\gets$ in terms of the right propagating (``incident") probe operator $a_p^\to$ and cavity operator $a_c^\to$.    

Starting with boundary condition (\ref{bclTeq}), we have $a_c^\gets (\omega, t_0)=-e^{2i\omega l/v_c} a_c^\to (\omega, t_0)$, so that the cavity phase field solution (\ref{phasepsolneq}) (with $i=c$) becomes 
\begin{eqnarray}
&&\phi_c (x,t)=\frac{2\pi}{\Phi_0}\int_0^{\infty} d\omega  \sqrt{\frac{\hbar Z_c}{\pi \omega}}\frac{1}{2}e^{-i\omega(t-t_0)}\cr
&&\times\left[e^{i\omega x/v_c} -e^{-i\omega (x-2l)/v_c}\right] a^{\to}_c (\omega,t_0) +{\mathrm{h.c.}};
\label{phasec2solneq}    
\end{eqnarray}
one may readily verify that solution (\ref{phasec2solneq}) vanishes at $x=l$ as required by the boundary condition (\ref{bclTeq}).
Using Eq. (\ref{phasepsolneq}) (for $i=p$), Eq. (\ref{phasec2solneq}), and  boundary condition (\ref{cpjunctioneq}), we can now couple the cavity and probe phase field to arrive at the following respective expressions for $\phi_p$ and $a_c^\to$:
\begin{eqnarray} 
&&\phi_p (x,t)=\frac{2\pi}{\Phi_0}\int_0^{\infty} d\omega  \sqrt{\frac{\hbar Z_p}{\pi\omega}}\frac{1}{2}e^{-i\omega t}\cr
&&\times\left[e^{i\omega x/v_p}+\left(\frac{1+i \omega  Z_p C_{pc}}{1-i\omega Z_p C_{pc}}\right) e^{-i\omega x/v_p}\right] a^{\mathrm{in}}_p (\omega)\cr
&&-i\frac{2\pi}{\Phi_0}\int_0^{\infty} d\omega  \sqrt{\frac{\hbar Z_p}{\pi\omega}}\frac{1}{2} e^{-i\omega(t-t_0+x/v_p)} \left(1-e^{2i\omega l/v_c}\right)\cr
&&\times\frac{\omega \sqrt{Z_p Z_c} C_{pc} }{1-i\omega Z_p C_{pc}} a_c^\to (\omega,t_0)  +{\mathrm{h.c.}}
\label{pumpphasepsoln3eq}
\end{eqnarray}
and
\begin{eqnarray}
\label{cavityasolneq}
&&\left[\cos\left(\omega l/v_c\right) -\frac{\omega  Z_c C_{pc}}{1+\left(\omega Z_p C_{pc}\right)^2} \sin\left(\omega l/v_c\right)\right]a_c^\to(\omega,t_0)\cr
&&-i\frac{\left(\omega \sqrt{Z_c Z_p} C_{pc}\right)^2 }{1+\left(\omega Z_p C_{pc}\right)^2} \sin\left(\omega l/v_c\right) a_c^\to(\omega,t_0)\cr
&&=-ie^{-i\omega \left(t_0+ l/v_c\right)} \frac{\omega\sqrt{Z_p Z_c}  C_{pc} }{1-i\omega Z_p C_{pc}} a^{\mathrm{in}}_p (\omega),
\label{pumpphasesolneq}
\end{eqnarray}
where $a^{\mathrm{in}}_p (\omega) \equiv e^{i\omega t_0} a_p^\to (\omega,t_0)$ may be interpreted classically as the right propagating component of the pump/probe line field in frequency space that enters the cavity at time $t=0$.

Under the condition of weak cavity-probe coupling, Eq. (\ref{pumpphasesolneq}) describes the Fourier transform of the quantum dynamics of approximately independent harmonic oscillators (i.e., cavity modes) subject to damping and noise. The resonant mode frequencies are obtained by setting the real, square-bracketed coefficient in the first line to zero and solving for $\omega$, while the mode linewidths are given by the imaginary coefficient on the second line of Eq. (\ref{pumpphasesolneq}). The term involving $a^{\mathrm{in}}_p (\omega)$ represents the pump drive and noise. In particular, imposing the condition of weak coupling given by the smallness of the dimensionless parameter $\xi \equiv C_{pc}/({\mathcal{C}}_c l)\ll 1$, and expanding to first order in $\xi$, we obtain for the mode frequencies
\begin{equation}
\label{bare_cavity_w0_field}
\omega_{n}\approx\left(2n+1\right)\frac{\pi v_c}{2l} \left(1-\frac{C_{pc}}{{\mathcal{C}}_c l}\right),\, n=0, 1, 2,\dots,
\end{equation}
which coincides with the lumped element expression (\ref{bare_cavity_w0_circuit}) for the cavity mode capacitance: $C_{\text{cav}}= \mathcal{C}_{c} l/2$.
Furthermore, under the Markovian approximation, the pump/probe damping rate $\kappa_{\text{ext}}$ is given by
\begin{equation}
    \kappa_{\text{ext}} = 2Z_p\frac{C_{pc}^2}{{\mathcal{C}}_c l}\omega_n^2,
    \label{cavdampeqn}
\end{equation}
which matches Eq. (\ref{bare_cavity_qc}) near $\omega_n$ with the external quality factor $Q_{\text{ext}} \equiv \omega_n / \kappa_{\text{ext}}$.

We can now use these results to derive the standard quantum Langevin equation in the Fourier domain involving the familiar closed-system cavity mode Hamiltonian, along with the zero-point fluctuations of the cavity phase coordinate modes, the details of which are given in Appendix \ref{bare_cavity_appendix}.
%[Eq. (\ref{ftql2eq})],  along with the explicit expressions for pump/probe damping rate [Eq. (\ref{cavdampeqn})]. These results lead to the familiar closed-system Hamiltonian [Eq. (\ref{barecavity_ham})] of a bare cavity, the details of which are given in Appendix \ref{bare_cavity_appendix}.

\subsection{Output power}
Experiments on the device performance require measurements on the steady state response of the cavity, subject to a pump with frequency $\omega_p$ typically applied in the vicinity of the fundamental cavity resonance $\omega_0$ given by Eq. (\ref{bare_cavity_w0_field}) for $n=0$. In practice, this involves a classical input pump signal at room temperature, which is further attenuated at different stages to reach the sample placed at the cryogenic temperature ($\lesssim 30$ mK), for which the scale of thermal fluctuations $k_BT \ll \hbar \omega_0$. In the absence of driving, we consider the continuum of modes in the semi-infinite transmission line to be in a thermal state given by
\begin{equation}
    \rho_{\text{th}} = \frac{1}{Z} \sum_{\{n(\omega)\}=0}^{\infty} e^{-\beta H_p} \; |\{n(\omega)\} \rangle_p \langle \{n(\omega)\}|_p
\end{equation}
where $|\{n(\omega)\} \rangle_p$ is the transmission line Fock state, $Z = {\text{Tr}} \left( e^{-\beta H_p} \right)$ is the partition function, $\beta \equiv 1/(k_B T)$, and the transmission line  Hamiltonian takes the form
\begin{equation}
    {H}_p = \hbar \int_0^{\infty} d \omega \omega \left(a_p^{\mathrm{in}}(\omega)\right)^{\dag} a_p^{\mathrm{in}}(\omega),
\end{equation}
where we neglect the zero point energy term since it does not contribute to the measured quantities.

The presence of driving may be approximated by a
displaced thermal state for the pump/probe transmission line:
$\rho_{\alpha,\text{th}} = D[\alpha] \rho_{\text{th}} D[\alpha]^{\dagger}$,\citep{barnett1985thermofield}
%\begin{equation}
%|\alpha\rangle_p= D [\alpha] %|0\rangle_p,
%\label{incohstateq}
%\end{equation}
where $D[\alpha]$ is a displacement operator, which is defined as follows:
\begin{equation}
    D[\alpha] = \exp\left(\int d\omega \left[ \alpha(\omega) \left(a_p^{\mathrm{in}}(\omega)\right)^{\dag}- \alpha^*(\omega) a_p^{\mathrm{in}}(\omega)\right]\right),
\end{equation} 
with 
\begin{equation}
\alpha(\omega)= \sqrt{\frac{P_p^{\mathrm{in}}T_p^2}{\hbar}} \frac{e^{-(\omega-\omega_p)^2 T_p^2/2}}{\sqrt{\omega}} e^{i\theta_p}.
\label{alphapumpeq}
\end{equation}
Here, $P_p^{\mathrm{in}}$ is the average pump power and $\theta_p$ is the pump phase. The pump coherence time $T_p$ is assumed to be longer than all other characteristic timescales of the system so that the displacement wavelet is narrowly smeared about $\omega = \omega_p$ in this large $T_p$ limit.

 We can then extract the time averaged output power in the bandwidth $\Delta\omega$ centered at $\omega_p$ using
\begin{equation}
    P_p^{\mathrm{out}}(\omega_p, \Delta\omega)=\overline{\left\langle \left[I_p^{\mathrm{out}}(x,t|\omega_p,\Delta\omega)\right]^2\right\rangle}Z_p,
    \label{poutexp}
\end{equation}
where the output probe current is 
\begin{equation}
    I^{\text{out}}_p(x,t)= -\frac{\Phi_0}{2\pi  {\mathcal{L}}_p} \frac{\partial \phi^{\text{out}}_p(x,t)}{\partial x},
\end{equation}
and the output phase field is
\begin{eqnarray}
    \phi_p^{\text{out}} (x,t)=  \frac{2\pi}{\Phi_0}\int_0^{\infty} d\omega  \sqrt{\frac{\hbar Z_p}{4 \pi \omega}}\; e^{-i\omega(t +x/v_p)} a_p^{\text{out}}(\omega).\cr &&
\end{eqnarray}
Following a similar convention as for $a^{\mathrm{in}}_p (\omega)$ given above [just after Eq. (\ref{pumpphasesolneq})], we define $a^{\mathrm{out}}_p (\omega) \equiv e^{i\omega t_0} a_p^\gets (\omega,t_0)$. Within the bandwidth $\Delta \omega$ and to first order in the capacitance ratio $\xi=C_{pc}/({\mathcal{C}}_c l)$, we can deduce $a^{\mathrm{out}}_p (\omega)$ by identifying the left propagating (i.e., reflected) terms involving the exponential factor $e^{-i\omega(t-t_0+x/v_p)}$ in the coupled cavity-probe relation (\ref{pumpphasepsoln3eq}). In short, we have
\begin{equation}
    a_p^{\mathrm{out}}(\omega)= a_p^{\mathrm{in}}(\omega)-i\sqrt{\kappa_{\text{ext}}} a_n(\omega),
    \label{newaouteq}
\end{equation}
the standard input-output relation for the cavity in a reflection mode measurement, where we have used the explicit expression (\ref{cavdampeqn}) for the pump/probe damping rate $\kappa_{\mathrm{ext}}$, and where the cavity mode annihilation operator is defined as follows:
\begin{equation}
    a_n(\omega) \equiv \sqrt{\frac{2l}{v_c}}e^{i\omega t_0} a_c^\to (\omega,t_0).
\end{equation}

Substituting the quantum Langevin equation (\ref{ftql2eq}) into the input-output relation (\ref{newaouteq})  and using the definition (\ref{poutexp}) for $P_p^{\mathrm{out}}$, we obtain
\begin{eqnarray}
    &&P_p^{\mathrm{out}}(\omega_p,\Delta\omega) = \frac{\hbar \omega_p}{4 \pi}\int_{\omega_p - \Delta \omega/2}^{\omega_p + \Delta \omega/2} d \omega \; |r(\omega)|^2 \cr && \times\left(1 + 2 \langle \left(a_p^{\mathrm{in}}(\omega)\right)^{\dag} a_p^{\mathrm{in}}(\omega) \rangle \right),
    \label{outpower_barecav}
\end{eqnarray}
where the cavity reflection coefficient $r(\omega)$ is defined as
\begin{equation}
    r(\omega) = \frac{\omega - \omega_n - i \kappa_{\text{ext}}/2}{\omega - \omega_n + i \kappa_{\text{ext}}/2}.
\end{equation}
For $\omega_p=\omega_0$ and $\Delta \omega \ll \omega_0$, we obtain
\begin{equation}
P_p^{\mathrm{out}}(\omega_p,\Delta\omega)=P_p^{\mathrm{in}}+\frac{\hbar\omega_p}{2 \pi}\int_{\omega_p - \Delta \omega/2}^{\omega_p + \Delta \omega/2} d \omega \left( n_p (\omega) + \frac{1}{2} \right),
\label{outpower3eq}
\end{equation}
with the transmission line average thermal occupancy $n_p(\omega) = (e^{\beta \hbar \omega} - 1)^{-1}$ (which is small in the frequency bandwidth of interest at $T\lesssim 30~{\text{mK}}$).
Since we set $\kappa_{\text{int}}=0$, the pump microwaves are reflected without any absorption/emission as expected.

\section{\lowercase{c}CPT-transmission line dynamics}
\label{cCPT}
Having validated the bare cavity-probe transmission line dynamics using the operator scattering approach, we can now extend the same approach to the cCPT system shown in Fig. \ref{ccpt_scheme}. In this section we present a first-principles derivation of the cCPT dynamics. To better account for actual devices, we allow for asymmetry in the Josephson junctions (JJs), given by distinct junction capacitances $C_{J1}$ and $C_{J2}$, and critical currents $I_{C1}$ and $I_{C2}$.

We begin by following the same procedure as in the previous section; in particular, we write down the cCPT-transmission line boundary conditions, which now accommodate the current through the CPT at $x=0$ (Fig. \ref{ccpt_scheme}). This leads to two additional phase degrees of freedom, one for each of the two JJs making up the CPT.  The CPT-cavity coupling accomplished through the flux biased SQUID loop reduces the number of independent phase coordinates from three down to two. We then proceed to write down the CPT Hamiltonian, and further use adiabatic elimination of the CPT dynamics to expand the resulting cavity effective potential about a stable minimum. The details of the operator scattering-based derivation of these results are given in Appendix \ref{ccpt_appendix}.
%The resulting anharmonic contributions from the CPT are highly tunable and very strong; their effects can be observed even close to the single photon limit. 
\subsection{Formulation of the circuit equations}
Referring to Fig. \ref{scheme}b, the cCPT consists of two JJs in series located at the voltage anti-node of the cavity, with the electrostatic energy of the CPT island tuned via a gate voltage $V_g$. The relevant coordinates for the cCPT system are the cavity phase field $\phi_c(x,t)$ and the JJ phase coordinates $\varphi_{1(2)}$. Note that Eqs. (\ref{phieq}), (\ref{bclTeq}) and (\ref{phasepsolneq}) remain the same, while the boundary condition (\ref{cpjunctioneq}) at $x=0^+$ gets modified to
\begin{eqnarray}
&&-\frac{\Phi_0}{2\pi{\mathcal{L}_p}}\left.\phi_p'(x,t)\right|_{x=0^-} =  C_{pc}\left.\left(\ddot{\phi}_c-\ddot{\phi}_p\right)\right|_{x=0} \nonumber \\[1pt] &=& \; -\frac{\Phi_0}{2\pi{\mathcal{L}_c}}\left.\phi_c'(x,t)\right|_{x=0^+} +\frac{\Phi_0}{2\pi}  C_{J1} \ddot{\varphi_1}  +I_{C1} \sin\varphi_1 \nonumber \\[1pt]
%\end{eqnarray}
%\begin{eqnarray}
%-&&\frac{\Phi_0}{2\pi{\mathcal{L}_p}}\left.\frac{\partial\phi_p}{\partial x}\right|_{x=0^-}%
&=& -\frac{\Phi_0}{2\pi{\mathcal{L}_c}}\left.\phi_c'(x,t)\right|_{x=0^+} + \frac{\Phi_0}{2\pi}  \left(C_{J2}+C_g\right) \ddot{\varphi_2} \cr &&+ I_{C2} \sin\varphi_2-C_g\dot{V_g}(t),
\label{phip2eq}
\end{eqnarray}
where $f'(x,t)$ and $\dot{f}(x,t)$ represent the spatial and temporal derivatives, respectively, and recall $\Phi_0 = h/2e$ is the flux quantum.

The associated SQUID loop constrains the phase coordinates through the relation
\begin{equation}
\varphi_1(t)+\varphi_2(t)-\phi_c(0,t)\approx 2\pi n +2\pi \frac{\Phi_{\mathrm{ext}}(t)}{\Phi_0},
\label{phaseTbceq}
\end{equation}
where $\Phi_{\mathrm{ext}}(t)$ is the externally applied flux bias, $n$ is an arbitrary integer (set to zero without loss of generality). For our cCPT device,\citep{brock-ccpt-2020} the magnitude of the supercurrent $I_{\mathrm{cir}}$ circulating through the cCPT loop %much less than the JJs' critical current,
is such that we can neglect the resulting induced flux, i.e., $ {( \mathcal{L}}_c l) I_{\mathrm{cir}} \ll \Phi_0$.
Equation (\ref{phaseTbceq}) allows us to reduce the number of system coordinates by one, since the average CPT coordinate $\bar{\varphi} = (\varphi_1+\varphi_2)/2$ determines the cavity phase $\phi_c(x,t)$; we will  utilize the cavity coordinate $\phi_c(x,t)$ and the half-difference CPT coordinate $\delta\varphi=(\varphi_1-\varphi_2)/2$ as the primary, independent variables. The equation of motion for $\delta \varphi$ can be obtained using the modified Eq. (\ref{phip2eq}) together with Eqs. (\ref{phasepsolneq}) (for $i=p$) and Eq. (\ref{phasec2solneq}). As we are primarily interested in deriving the charge sensitivity of the device in the present work, we only allow a time dependent gate voltage modulation and neglect any time dependent magnetic flux modulation. We obtain:
\begin{eqnarray}
 \frac{\Phi_0 C_{\text{CPT}} C_{\Sigma}}{\pi} \ddot{\delta \varphi} = && \;(C_g +\Delta C_J) C_{pc} \frac{\partial \hat{V}^{\mathrm{in}}_p(0,t)}{\partial t} \cr &&  -({\mathrm{cCPT}}~{\mathrm{terms}}) ,
 \label{differencenoiseeq}
\end{eqnarray}
where the junction capacitance asymmetry $\Delta C_J = C_{J2} - C_{J1}$, the CPT capacitance $C_{\text{CPT}} = C_{J1} (C_{J2} + C_g)/C_{\Sigma}$ and the total island capacitance $C_{\Sigma} = C_{J1}+C_{J2}+C_g$.

The `cCPT terms' contribution in Eq. (\ref{differencenoiseeq}) is given by
\begin{eqnarray}
    && \left[I_{C1} \left(C_g +C_{J2}\right)-I_{C2} C_{J1}\right] \sin\left(\phi/2 \right) \cos\left(\delta\varphi\right) \nonumber \\[2pt]
&+& \left[I_{C1} \left(C_g +C_{J2}\right)+I_{C2} C_{J1}\right] \cos\left(\phi/2 \right) \sin\left(\delta\varphi\right) \nonumber \\[2pt] &+& C_g C_{J1} \dot{V}_g,
\label{cCPTtermseq}
\end{eqnarray}
where we have introduced a displaced  cavity phase $\phi(t)$ to absorb the external flux bias as follows: 
\begin{equation}
\label{displaced_cavity_phase_eqn}
    \phi(t) = \phi_c(0,t)+2 \pi\Phi_{\mathrm{ext}}/\Phi_0.
\end{equation}
The first term on the RHS of Eq. (\ref{differencenoiseeq}) represents the CPT's direct coupling to the pump/probe line:
\begin{eqnarray}
    &&\hat{V}^{\mathrm{in}}_p(0,t)=  -i \int_0^{\infty} d\omega  \sqrt{\frac{\hbar\omega Z_p}{\pi}}e^{-i\omega t} \cr &&
    \times\left(1-i \omega Z_p C_{pc}\right)^{-1} a_p^{\mathrm{in}}(\omega) +{\mathrm{h.c}},
    \label{translineV2eq}
\end{eqnarray}
to be contrasted with  the more familiar  indirect  CPT coupling to the probe line via the cavity. As we will see in the next steps, the former contribution appears as an unwanted gate modulation, which can however be neglected as long as $C_g \ll C_J$.

We may now similarly proceed as in Sec. \ref{fullanalysis} to employ the equation of motion for the cavity phase $\phi_c(x,t)$, and further determine the Lagrangian  and Hamiltonian of the cCPT system. However, as this turns out to be a cumbersome task if no approximations are made, we will first focus on the half-difference CPT coordinate $\delta\varphi$, utilizing several valid approximations to simplify the analysis. 

\subsection{Adiabatic elimination of CPT dynamics}
Instead of writing down the open cCPT Hamiltonian which contains contributions from the cavity, CPT, and the pump/probe transmission line, we use Eq. (\ref{differencenoiseeq}) to first obtain the CPT contribution to the Lagrangian, which then yields the following CPT Hamiltonian:
\begin{eqnarray}
&&H_{\text{CPT}}=\nonumber \\[1pt]
&&\left(\frac{2\pi}{\Phi_0}\right)^2 \frac{1}{8 C_{\text{CPT}}}
\left( p_{\delta\varphi} - \frac{\Phi_0}{\pi}   \left(\frac{C_{J1}}{C_{\Sigma}} C_g V_g - \hat{Q}_p^{\text{in}}(0,t) \right) \right)^2 \nonumber \\[1pt]
&&- 2  E_J \cos\left(\phi/2\right) \cos\left(\delta\varphi\right) + 2 \delta E_J \sin\left(\phi/2\right) \sin\left(\delta\varphi\right),\nonumber \\
\label{hameq}
\end{eqnarray}
where $p_{\delta\varphi}$ is the momentum conjugate to the half-difference CPT phase coordinate $\delta\varphi$, $\hat{Q}_p^{\text{in}}(0,t) =\left(C_g+ \Delta C_{J}\right) C_{pc} \hat{V}^{\mathrm{in}}_p(0,t)/C_\Sigma$, and 
the effective Josephson energy coefficients in the potential energy term are defined as follows:
\begin{equation}
    E_J = \left(E_{J1}+\frac{ C_{J1}}{C_{\Sigma}}\Delta E_J \right)
\end{equation}
and
\begin{equation}
    \delta E_J = \frac{\left[\left(C_g+ C_{J2} \right) E_{J1}- E_{J2} C_{J1}\right]}{C_{\Sigma}},
\end{equation}
 with the Josephson energies of the junctions defined as $E_{J1(2)} = I_{C 1(2)}\Phi_0/2\pi$ and $\Delta E_J = E_{J2} - E_{J1}$.
 
\begin{figure*}[thb]
%\subfloat{\label{state_approx}\includegraphics[width=0.3\textwidth,trim=0 0 0 0, clip]{state_approx.eps}}
\subfloat{\label{cpt_energy_levels} \includegraphics[width=0.45 \textwidth,trim=0 500 450 0, clip] {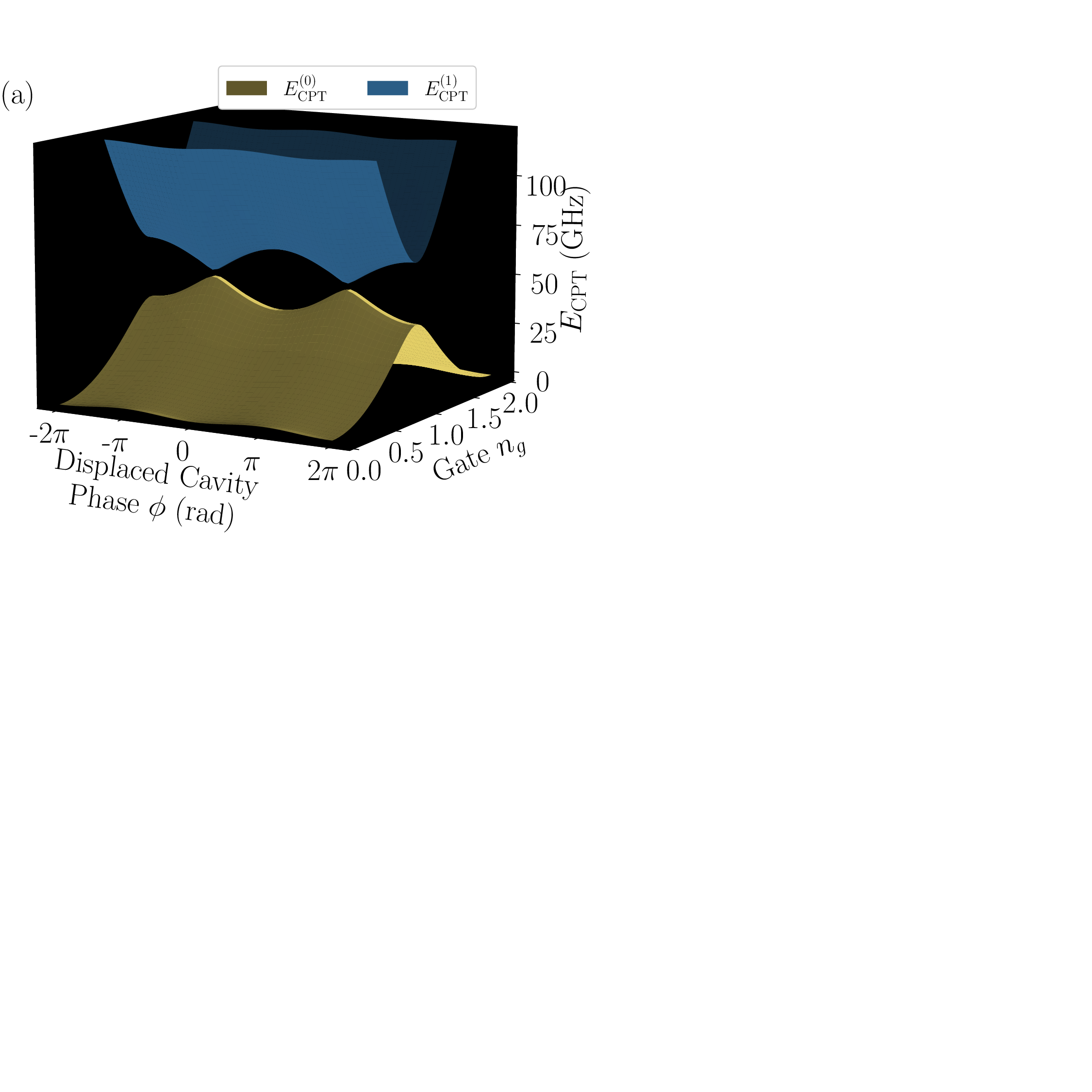}}
\subfloat{\label{cpt_energy_splitting}
\includegraphics[width=0.4 \textwidth,trim=0 0 0 0, clip] {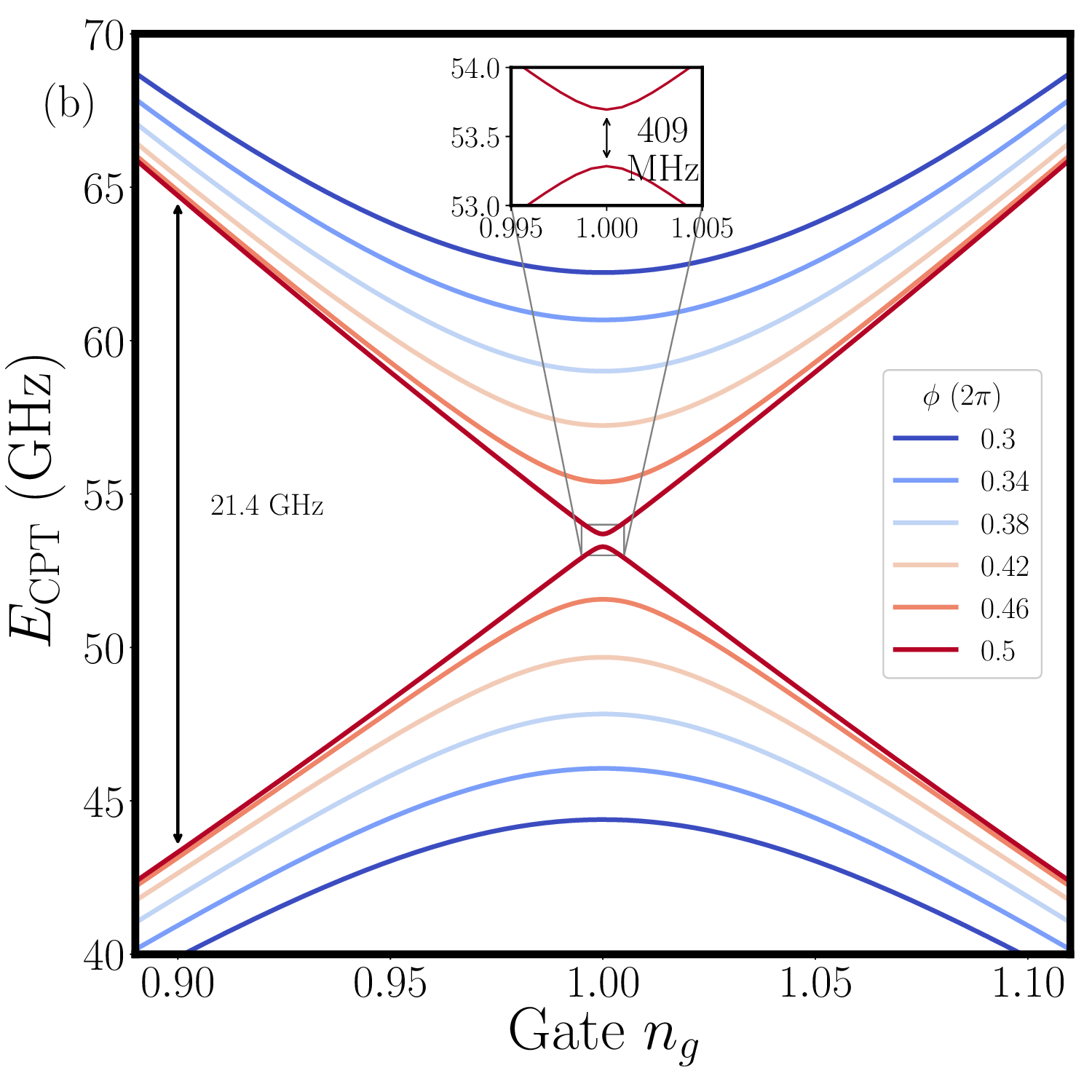}}
\caption{(a) %Maximum absolute error in the CPT ground state energy for different charge state number-truncations  as a function of the displaced cavity phase $\phi$. The maximum value is determined by scanning across the entire range of gate-bias $n_g$. (b)
The ground and first excited energy band-structure of the CPT. Note that the adiabatic approximation may break down in the vicinity of charge degeneracy: $n_g=\pm 1$. (c) Energy splitting between the ground and first excited state in the vicinity of charge degeneracy. For $|1-n_g| \geq 0.1$, the adiabatic approximation holds since the energy splitting is much greater than the characteristic frequencies of the system. The parameter values used for these simulations are provided in Table \ref{parameter_table}.}\label{cpt_energy_charac}
\end{figure*}

The corresponding quantized CPT operators obey the commutation relations $[\hat{\delta\varphi},\hat{N}]=i$, where $\hat{N} \equiv \hat{p}_{\delta\varphi}/\hbar$. In the more suitable phase coordinate form with unit circle configuration space, the commutation relations take the form (neglecting hats):
\begin{equation}
\left[{e^{i{\delta\varphi}}},{N}\right]=-{e^{i{\delta\varphi}}}.
\label{expphasecreq}
\end{equation}
Equation (\ref{expphasecreq}) has a Hilbert space representation spanned by the eigenstates $|N\rangle$ of the operator $\hat{N}$:
\begin{equation}
\hat{N}|N\rangle=N |N\rangle,\, N=0,\pm 1,\pm 2,\dots
\label{Nopeq}
\end{equation}
i.e., $N$ takes discrete, integer values which can be interpreted as the number of excess Cooper pairs on the CPT island. Similarly, we can also define the gate polarization number $n_g$ in single electron units as follows:
\begin{equation}
n_g \equiv \frac{2 \Phi_0 C_{J1} C_g}{\hbar\pi C_{\Sigma}}  V_g=\frac{2 C_{J1} C_g}{e C_{\Sigma}} V_g
.\label{inducedchargeeq}
\end{equation}

The CPT Hamiltonian then becomes
\begin{eqnarray}
&&H_{\mathrm{CPT}} = 4 E_C \sum_{N=-\infty}^{+\infty} \left[N-\frac{1}{2}\left({{n_g}-\hat{N}^{\mathrm{in}}_p}\right)\right]^2 |N\rangle\langle N|\cr
&&-E_J  \cos \left(\phi/2 \right) \sum_{N=-\infty}^{+\infty}\left( |N+1\rangle\langle N| +|N-1\rangle\langle N|\right)\cr
&&-i \delta E_J \sin \left(\phi/2\right)\sum_{N=-\infty}^{+\infty}\left( |N+1\rangle\langle N| -|N-1\rangle\langle N|\right),\cr
&&
\label{dressedhamneq}
\end{eqnarray}
where $\phi$ is defined in Eq. (\ref{displaced_cavity_phase_eqn}), the charging energy $E_C =e^2/(8 C_{\text{CPT}})$, and the effective, polarization charge number noise operator is given by
$\hat{N}^{\mathrm{in}}_p(t)=2 \hat{Q}_p^{\text{in}}(0,t)/e$. Equation (\ref{dressedhamneq}) reduces to the familiar form of the CPT Hamiltonian in the limiting case of junction symmetry $\Delta E_J = \Delta C_J = 0$ and $C_g \ll C_J$ (with $C_J\equiv C_{J1}=C_{J2}$):\citep{joyez-single-nodate}
\begin{eqnarray}
&&H_{\mathrm{CPT}} = 4E_c \sum_{N=-\infty}^{+\infty} \left(N-\frac{n_g}{2}\right)^2 |N\rangle\langle N|\cr
&&- E_J\cos(\phi/2) \sum_{N=-\infty}^{+\infty}\left( |N+1\rangle\langle N| +|N-1\rangle\langle N|\right),\cr
&&
\end{eqnarray}
where $E_C\approx e^2/{(2\cdot 2 C_J)}$.

\begin{table*}
\caption{\label{parameter_table} Numerical values of the parameters used in the simulations. The parameters are based on the experimental cCPT device.\cite{brock-ccpt-2020}}
\scalebox{1}{
\footnotesize
\begin{tabular} {|l|r|}
\hline
Parameter & Value \\
\hline
Length of microwave resonator $l$ & 5135 $\mu$m \\
Capacitance per unit length $\mathcal{C}_c$ & 0.17 nF/m \\
Inductance per unit length $\mathcal{L}_c$ & 0.41 $\mu$H/m \\
Coupling capacitance $C_{pc}$ & 7.95 fF\\
Bare cavity resonance $\omega_0$ & 5.76 GHz\\
\hline
\end{tabular}
\hspace{5pt}
\begin{tabular} {|l|r|}
\hline
Parameter & Value \\
\hline
CPT capacitance $C_{\mathrm{CPT}}$ & 90 aF\\
Gate capacitance $C_g$ & 6.27 aF\\
Charging energy $E_C/h$ & 53.49 GHz\\
Josephson energy $E_J/h$ & 15.17 GHz\\
Asymmetry in Josephson energy $\delta E_J$ & 205 MHz \\
\hline
\end{tabular}}
\end{table*}
Treating $\phi(t)$ and $N_p^{\mathrm{in}}$ as static, commuting numbers, the Hamiltonian (\ref{dressedhamneq}) can be diagonalized assuming an approximate, finite dimensional Hilbert space  truncation to obtain the CPT energy eigenvalues. Figure \ref{cpt_energy_charac} shows the CPT ground and first excited energy eigenvalue characteristics within a gate polarization range $0 \leq n_g \leq 2$ and a displaced cavity phase range $0 \leq \phi \leq 2\pi$. 
Note that the assumed parameter values used in our simulations take into account a small asymmetry in the JJ energies (See Table \ref{parameter_table}). %Figure  \ref{state_approx} plots 
As the maximum error of the CPT ground energy as a function of $\phi$ is negligible for a five charge state approximation relative  to a ten charge state basis truncation, we employ the five charge state basis for our simulations.

Assuming small $N_p^{\text{in}}$, we see that the CPT approaches charge degeneracy as $n_g \to \pm 1$ (Fig. \ref{cpt_energy_levels}). As a result, the system has an increased probability of transitioning to the first excited energy eigenstate in this limit. The experimental characterization\cite{brock-ccpt-2020} also observes quasiparticle poisoning close to charge degeneracy, as a consequence of lower electrostatic energies of odd electron-states as compared to the CPT charging energy.\citep{aumentado-nonequilibrium-2004,lutchyn-effect-2007} Taking into account both these factors, we further limit our considered gate polarization range to $-0.9 \leq n_g \leq 0.9$. The CPT level splitting between the ground and excited states over this modified range of bias space $(n_g,\phi)$ is much larger than the other characteristic frequencies of the system, namely the bare cavity fundamental mode frequency   and similar drive frequency ($\approx 5.76$ GHz), and the gate modulation frequency $\omega_g$ ($\sim$ tens of MHz) (Fig \ref{cpt_energy_splitting}). We thus impose the valid and essential approximation going forward to the effect that if the cavity `dressed' CPT is initially in its lowest energy eigenstate with energy $E_\text{{CPT}}^{(0)}$, it will remain in this state for the duration of the measurement, evolving adiabatically.
\subsection{Effective cavity Hamiltonian}
The adiabatic elimination of the CPT from the total Hamiltonian dynamics effectively replaces the Hamiltonian (\ref{dressedhamneq}) by its ground state energy $E_{\text{CPT}}^{(0)}$, which can subsequently be used to obtain the cavity phase equation of motion counterpart to Eq. (\ref{differencenoiseeq}). Invoking the wave equation (\ref{phieq}) and boundary condition (\ref{phip2eq}), we arrive at
\begin{eqnarray}
&& \phi_c'(0^+,t) - \frac{C_{\text{CPT}}}{{\mathcal{C}}_c }\phi_c''(0^+,t)-\frac{\mathcal{L}_c}{\mathcal{L}_p}\phi_p'(0^-,t) \cr && -\left(\frac{2\pi}{\Phi_0}\right)^2 {\mathcal{L}}_c \frac{\partial E_{\mathrm{CPT}}^{(0)}}{\partial \phi_c} = -\frac{2\pi {\mathcal{L}}_c C_g C_{J1}}{\Phi_0 C_{\Sigma}}\dot{V}_g.
\label{phi4ppeq}
\end{eqnarray}
 We identify the above expression as the modified boundary condition at $x=0$ coupling the cavity and pump/probe transmission line, and including the dressed CPT contribution as a perturbation [c.f. Eq.~(\ref{cpjunctioneq})]. We may now follow the same operator scattering method steps as carried out for the bare cavity case in Sec. \ref{bare} to obtain the renormalized resonant cavity fundamental frequency. Before deriving this explicitly, we first simplify Eq. (\ref{phi4ppeq}) by renormalizing the bare cavity Hamiltonian (\ref{barecavity_ham}), which now has an effective potential given by 
 \begin{equation}
     V_{\text{eff}} = \left(\Phi_0/2\pi\right)^2 \phi_0^2/2 L_{0} + E_{\text{CPT}}^{(0)} (\phi_0)
\label{effective_potential_expression}    
\end{equation}
restricted to the fundamental phase coordinate mode $\phi_0$, where $L_0 = 8 \mathcal{L}_c l/\pi^2$ is the corresponding fundamental mode inductance (see Appendix \ref{bare_cavity_appendix}). 

\begin{figure}[thb]
\centering
\subfloat{\label{pot_min_image} \includegraphics[width=0.23 \textwidth,trim=0 0 0 0, clip] {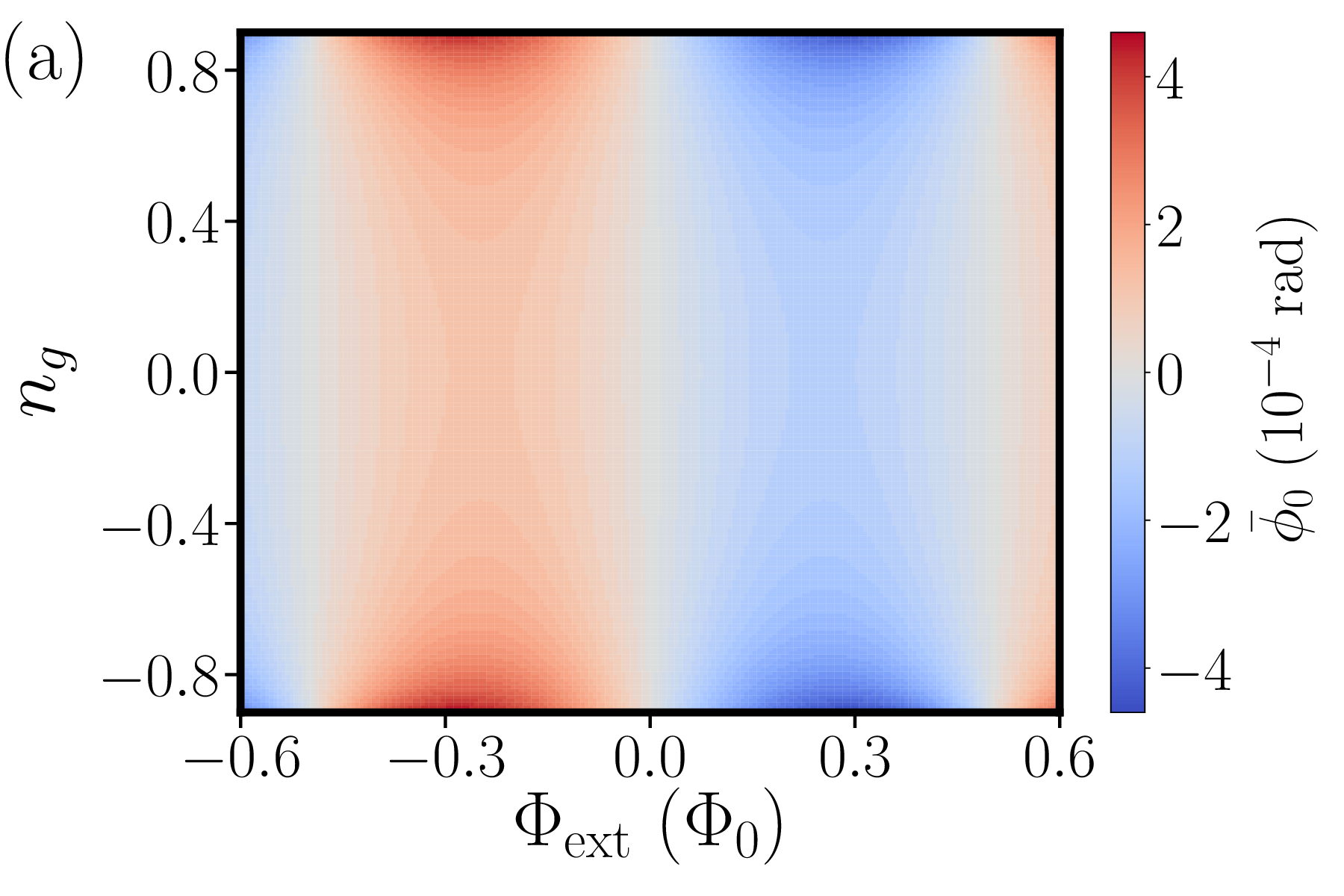}}
\subfloat{\label{lcptratio}\includegraphics[width=0.23\textwidth,trim=0 0 0 0, clip]{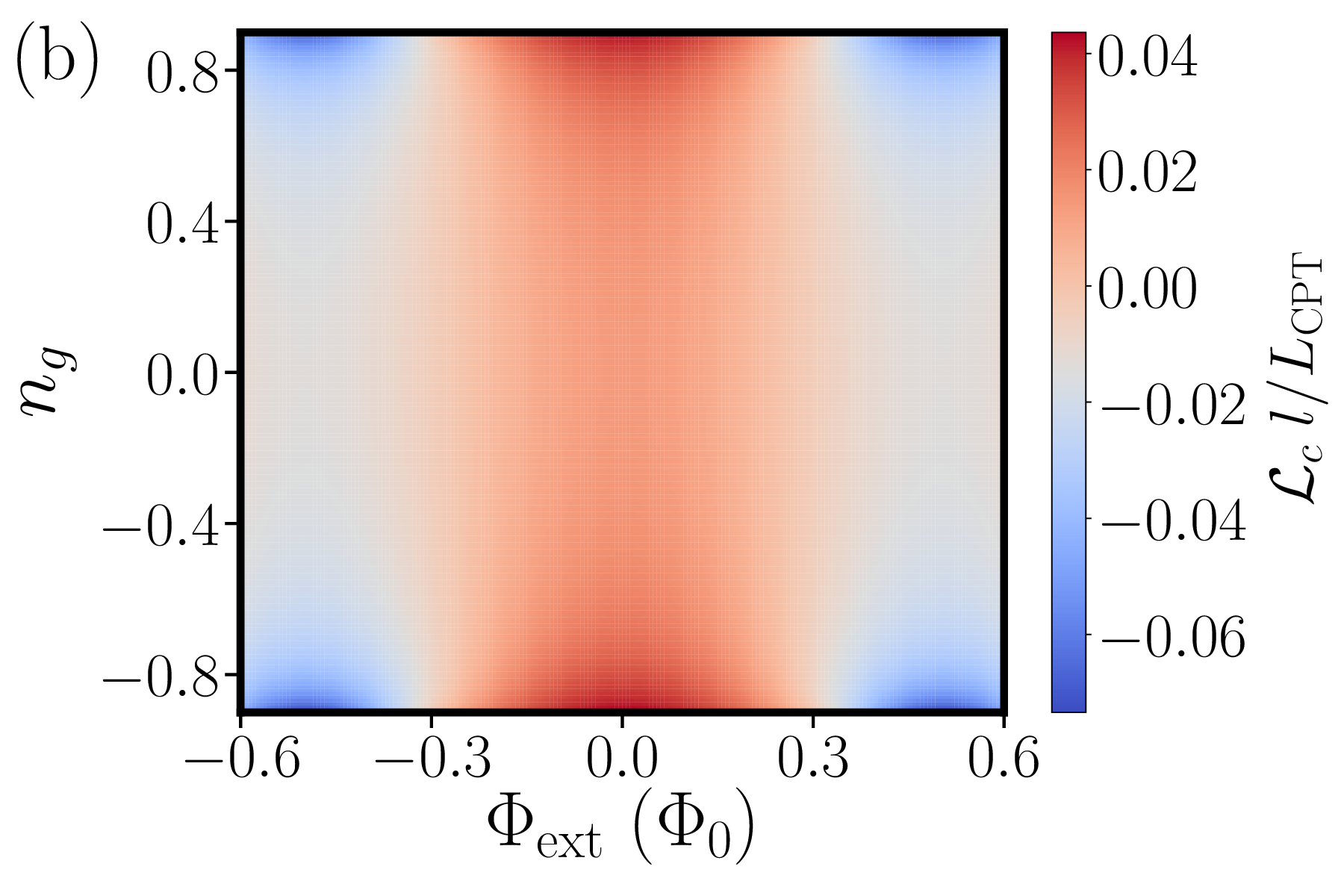}}\\
\subfloat{\label{resfreq_ccpt_image} \includegraphics[width=0.23 \textwidth,trim=0 0 0 0, clip] {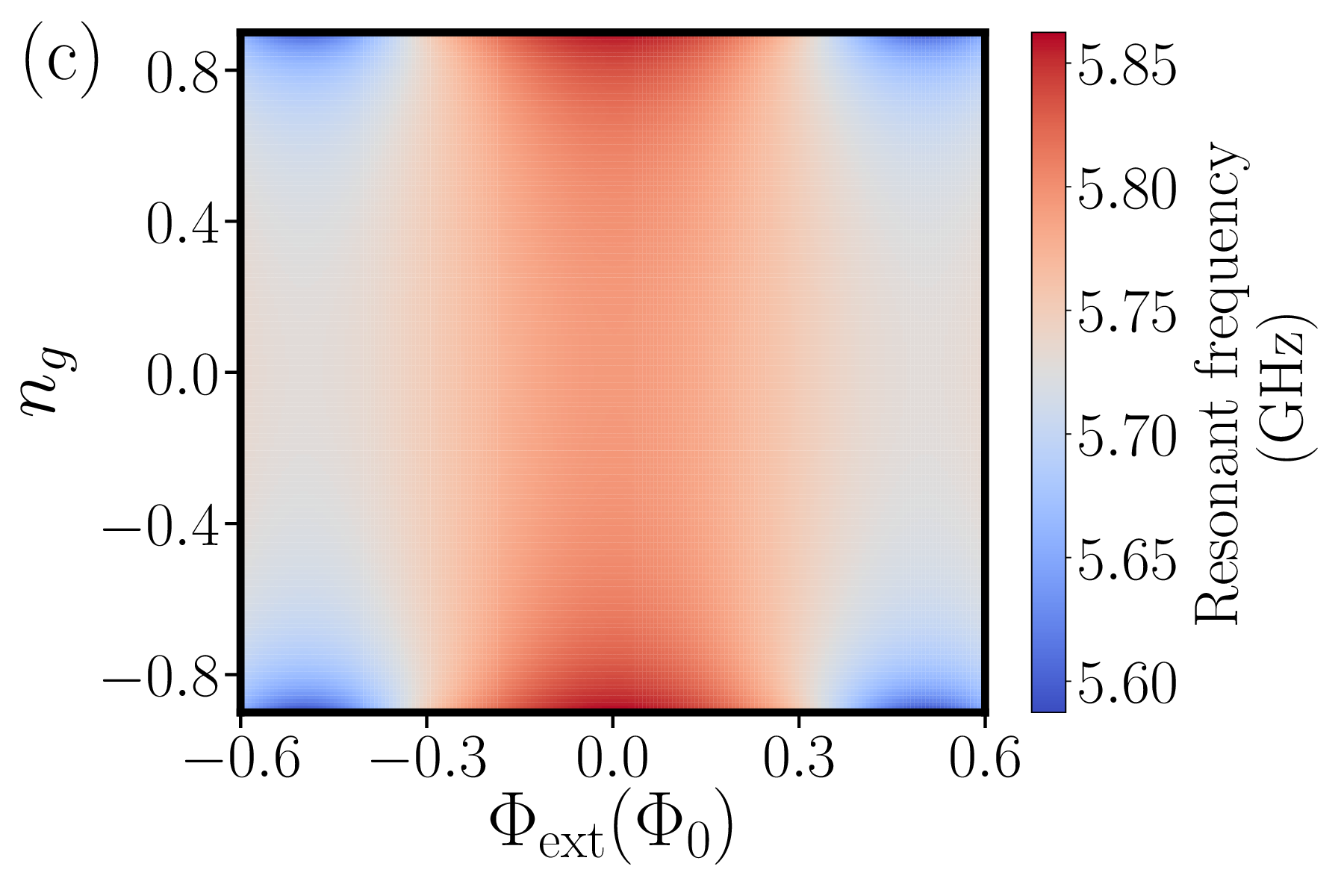}}
\subfloat{\label{zpf} \includegraphics[width=0.23 \textwidth,trim=0 0 0 0, clip] {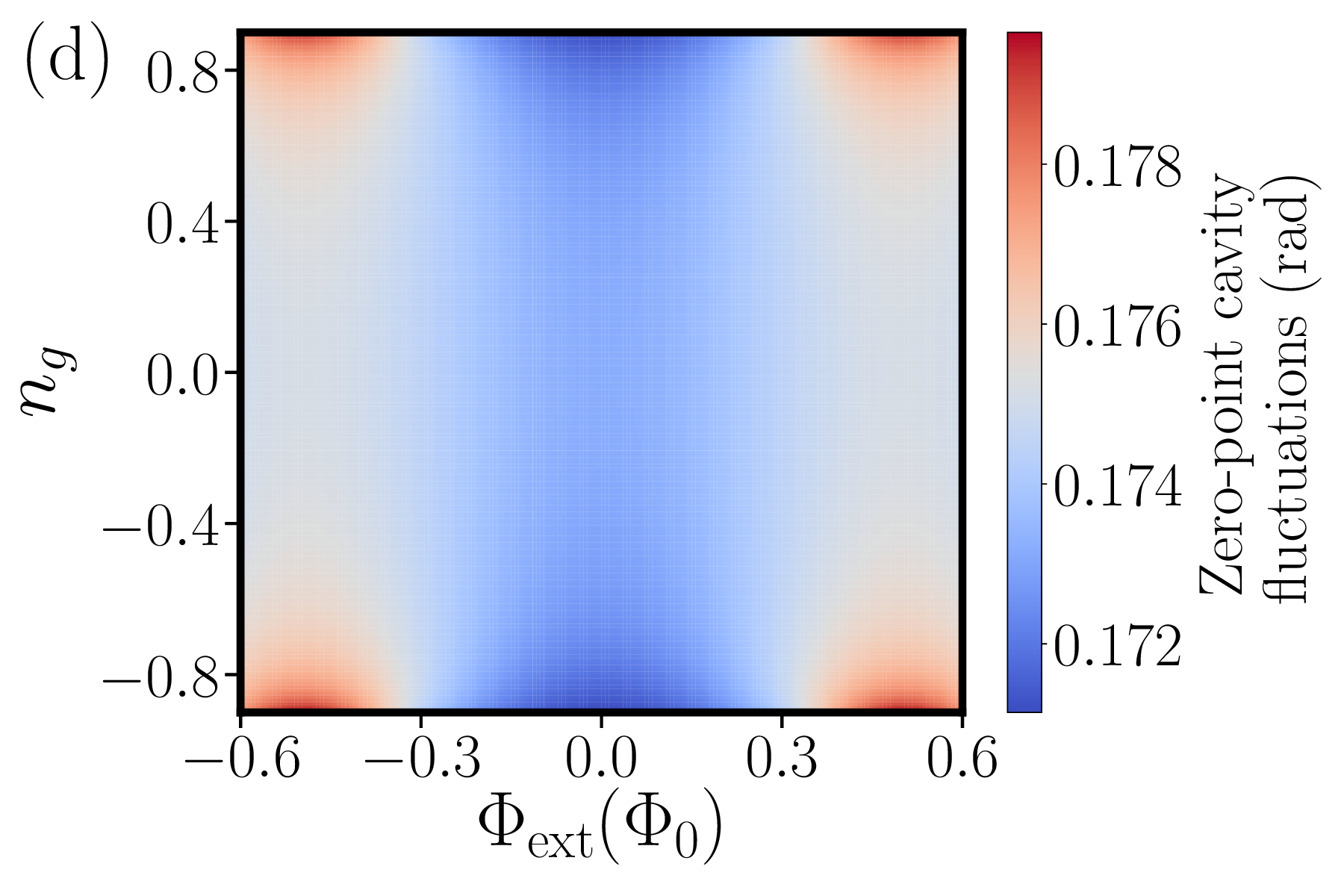}}
\caption{(a) The shift in the equilibrium point corresponding to the minimum effective potential energy as function of $n_g$ and $\Phi_{\text{ext}}$. (b) The smallness of the ratio of cavity inductance to the CPT inductance ensures that the CPT weakly perturbs the cavity.  (c) Resonance frequency shift of the cavity across the tunable bias range. (d) Plot of zero-point fluctuations as a function of $n_g$ and $\Phi_{\text{ext}}$. The shift from the original value is negligible for our system.} \label{ccpt_charact}
\end{figure}

The CPT introduces anharmonicity to varying orders when expanded about the equilibrium point $\bar{\phi}_0 (n_g, \Phi_{\text{ext}})$ obtained through the condition,
\begin{equation}
\left( \left. \left(\frac{\Phi_0}{2\pi}\right)^2 \frac{\phi_0}{L_0} + \frac{\partial E_{\mathrm{CPT}}^{(0)}}{\partial \phi_0}\right)\right|_{\bar{\phi}_0}=0.
\label{potmin3eq}
\end{equation}
As is evident in Fig \ref{pot_min_image}, this shift in equilibrium is much less than one in magnitude over the considered $(n_g, \Phi_{\text{ext}})$ bias range, and can be neglected. 
This simplifies the dependence $\phi (\phi_c, \Phi_{\text{ext}})$ to $\phi (\Phi_{\text{ext}})$ in Eq. (\ref{displaced_cavity_phase_eqn}).

In the limit where the CPT weakly perturbs the cavity fundamental resonance, i.e., ${C_{\mathrm{CPT}}}/{{\mathcal{C}}_c l}$ and ${{\mathcal{L}}_c l}/{L_{\mathrm{CPT}}} \ll 1$ (Fig \ref{lcptratio}), we obtain for the renormalized resonance frequency of the cCPT system coupled to the probe transmission line (see Appendix \ref{ccpt_appendix} for the detailed derivation):
\begin{equation}
    \omega_0(n_g,\Phi_{\text{ext}}) \approx \frac{\pi v_c}{2l}\left[1-\frac{C_{pc}+C_{\mathrm{CPT}}}{{\mathcal{C}}_c l}+\left(\frac{2}{\pi}\right)^2\frac{{\mathcal{L}}_c l}{L_{\mathrm{CPT}}}\right].
    \label{rencavfreqeq}
\end{equation}
where the CPT inductance $L_{\text{CPT}}$ can be defined from the curvature of $E_{\text{CPT}}$ as
\begin{eqnarray}
    %L_{\mathrm{CPT}}^{-1}= \left(\frac{2\pi}{\Phi_0}\right)^2\left.\frac{\partial^2 E_{\mathrm{CPT}}^{(0)}}{\partial \phi^2}\right|_{(n_g,2 \pi \Phi_{\text{ext}}/\Phi_0)}.
    L_{\mathrm{CPT}}^{-1}= \left(\frac{2\pi}{\Phi_0}\right)^2 \left.\frac{\partial^2 E_{\mathrm{CPT}}^{(0)}}{\partial \phi_{c}^2} \right|_{\phi_c=0} = \frac{\partial^2 E_{\mathrm{CPT}}^{(0)}}{\partial \Phi_{\text{ext}}^2}
    \label{CPTLeq_main}
\end{eqnarray}
The lumped element expression (\ref{ccpt_res_circuit}) with $C_{\text{cav}}={\mathcal{C}}_c l/2$ and $L_{\text{cav}}=8 {\mathcal{L}}_c l/\pi^2$ coincides with Eq. (\ref{rencavfreqeq}).
 
As mentioned above, the higher terms in the expansion of the effective potential $V_{\text{eff}}$ give rise to anharmonicity in the combined cCPT system which takes the form:
\begin{eqnarray}
&&V_{\text{eff}} = \left(\frac{\Phi_0}{2\pi}\right)^2 \frac{\phi_0^2}{2L_0} \cr && + \sum_{n=2}^{\infty} \; \sum_{k=2}^n \frac{1}{n!} {n\choose k} \phi_0^{k} \delta {n}_g^{n-k}
\left.\frac{\partial^{n} E_{\mathrm{CPT}}^{(0)}}{\partial \phi_0^{k} \partial n_g^{n-k}}\right|_{n_g=n_g^{(0)},\phi_0=0},\cr
&&\label{vexpeq}
\end{eqnarray}
where renormalization and having the minimum potential at $\phi_0 \sim 0$ lead to vanishing terms for $k=0$ and 1,  respectively. Expression (\ref{vexpeq}) also involves an expansion in the gate polarization variation $\delta n_g$ in order to account for gate voltage modulations relevant for electrometry (discussed in Sec. \ref{chargesens}). The total Hamiltonian is
\begin{eqnarray}
H_{\text{cCPT}} &=& \left(\frac{2\pi}{\Phi_0}\right)^2 \frac{p_{0}^2}{2 C_{0}} + V_{\text{eff}},
\end{eqnarray}
where $C_0$ is renormalized to $C_0 + C_{\text{CPT}}$ following the renormalized frequency expression in Eq. (\ref{rencavfreqeq}). As for the bare cavity case (see Appendix \ref{bare_cavity_appendix}), the phase operator of the fundamental cavity mode is expressed in terms of the photon creation/annihilation operators as follows:
$\phi_0 = \phi_{zp} \left(a_0 + a_0^{\dagger}\right)$,
with the zero-point fluctuations given by [c.f. eq. (\ref{zpteq})]
\begin{equation}
    \phi_{zp} = \left( \frac{2 \pi} {\Phi_0} \right) \sqrt{\frac{\hbar} {2 C_0 \omega_0}}.
\end{equation}
The generalized nonlinear cCPT Hamiltonian thus becomes
\begin{equation}
    H_{\text{cCPT}} =  \hbar \omega_0(n_g,\Phi_{\text{ext}}) a_0^{\dagger} a_0 + \sum_{n=3}^{\infty} \sum_{k=2}^n V_{n,k} \left(a_0 + a_0^{\dagger}\right)^k,
    \label{effcHam}
\end{equation}
where
\begin{equation}
    V_{n,k} = \frac{1}{n!} {n\choose k} \phi_{zp}^k \delta n_g^{n-k}
\left.\frac{\partial^{n} E_{\mathrm{CPT}}^{(0)}}{\partial \phi^{k} \partial n_g^{n-k}}\right|_{n_g^{(0)},2 \pi\Phi_{\mathrm{ext}}/\Phi_0}.
\label{potential_ten}
\end{equation}

We now make a few remarks about the Hamiltonian (\ref{effcHam}).
First, the tunability of the cavity frequency results in the tunability of the zero-point fluctuations of the cavity phase coordinate itself, i.e., $\phi_{zp} = \phi_{zp}(n_g,\Phi_{\text{ext}})$. Typical applications of similar devices generally operate in the high-photon limit, where the relatively small variations in the zero-point motion of the cavity do not have a substantial effect. In the low-photon limit however, the tunability in the zero-point fluctuations can become relevant, as this may potentially be utilized to access stronger quantum fluctuation regimes. For our device, the range of variation of $\phi_{zp}$ is found to be $\sim 5\%$ in the tunability range of our interest (Fig \ref{zpf}). 

Second, the experimental characterization is typically conducted in the limit of small gate modulation magnitude $|\delta n_g| \ll 1$. Additionally, the noise $N_p^{\text{in}}$ originating via the probe coupling to the CPT can also be neglected as long as $C_g \ll C_{\Sigma}$. We may thus restrict the potential energy expansion in (\ref{effcHam}) to first order in $\delta n_g$. 

Third, we may use a rotating wave approximation (RWA) to simplify the Hamiltonian to contain only terms leading to an unchanged photon number in the cavity. The validity of this approximation becomes explicit when we transform to the rotating frame of the pump frequency $\omega_p$, driven near the fundamental resonance $\omega_0$; contributions leading to changing photon number rapidly oscillate in this frame, and can thus be neglected. Consequently, we arrive at the simplified Hamiltonian of the cCPT device, valid up to $\mathcal{O}(\phi_0^2)$:
\begin{equation}
\label{final_ccpt_ham}
    H_{\text{cCPT}} =  \hbar \left(\omega_0 + g \delta n_g \right) a_0^{\dagger} a_0,
\end{equation}
%\begin{equation}
%\label{final_ccpt_ham}
 %   H_{\text{cCPT}} =  \hbar \left(\omega_0+ 12K  + g \delta n_g \right) a_0^{\dagger} a_0 + 6 \hbar K a_0^{\dagger2} a_0^2,
%\end{equation}
where the gate polarization coupling $g$ is given by
\begin{eqnarray}
    g = \frac{\phi_{zp}^2}{\hbar} \; \left.\frac{\partial^{3} E_{\mathrm{CPT}}^{(0)}}{\partial \phi^{2} \partial n_g}\right|_{(n_g^{(0)},2 \pi\Phi_{\mathrm{ext}}/\Phi_0)}.
    %K = \frac{\phi_{zp}^4}{24 \hbar} \; \left.\frac{\partial^{4} E_{\mathrm{CPT}}^{(0)}}{\partial \phi^{4}} \right|_{(n_g^{(0)},2 \pi\Phi_{\mathrm{ext}}/\Phi_0)},
\end{eqnarray}
The complete experimental characterization of the cCPT device following this theoretical model is given in Ref. \onlinecite{brock-ccpt-2020}.

\section{\lowercase{c}CPT as a linear electrometer}
\label{chargesens}
The highly tunable and strongly nonlinear nature of the cCPT is evident from the analysis in the previous section. In this section, we narrow our focus to examine the operation of the cCPT as a linear charge detector. A comprehensive understanding in the linear response regime is an essential first step before widening the scope of the device operation to include nonlinear contributions, for example to realize phase-sensitive amplification via squeezing.

In the simplest terms, we see from Eq. (\ref{final_ccpt_ham}) how a sinusoidal modulation in the gate charge appears as a renormalization-shift in the cavity resonance frequency. In particular, this gate modulation may be induced using a mechanical quantum dynamical system coupled at the CPT gate,\citep{rimberg-cavity-cooper-2014} thus facilitating sensing of the mechanical system via charge detection. A typical measurement involves driving the cavity near resonance, and detecting the sidebands via measurements of the output power averaged over some time $T_M$ that is long compared to the characteristic time-scales of the cCPT-mechanical system dynamics. 

In line with such a scheme, we will first look into the output power generation in the presence of an electrically simulated, sinusoidal gate modulation ``signal" $\delta n_g(t) = \delta n_g^{(0)} \cos\left(\omega_g t\right)$. This will enable a determination of the charge sensitivity of the cCPT in the low-average photon number drive limit, which we will find to be comparable to previously reported or predicted values for electrometers.\citep{roschier-noise-2004,sillanpaa-charge-2005,tosi-design-2019,zorin-quantum-limited-1996,brenning-ultrasensitive-2006,xue-measurement-2009,Zorin:2001} Most importantly, the behavior of the cCPT in this low drive power regime is limited by photon shot noise in the transmission line, which results in an attainable quantum-limited lower bound for charge sensitivity.

\begin{figure*}[thb]
\subfloat{\label{g_image} \includegraphics[width=0.4 \textwidth,trim=0 0 0 0, clip] {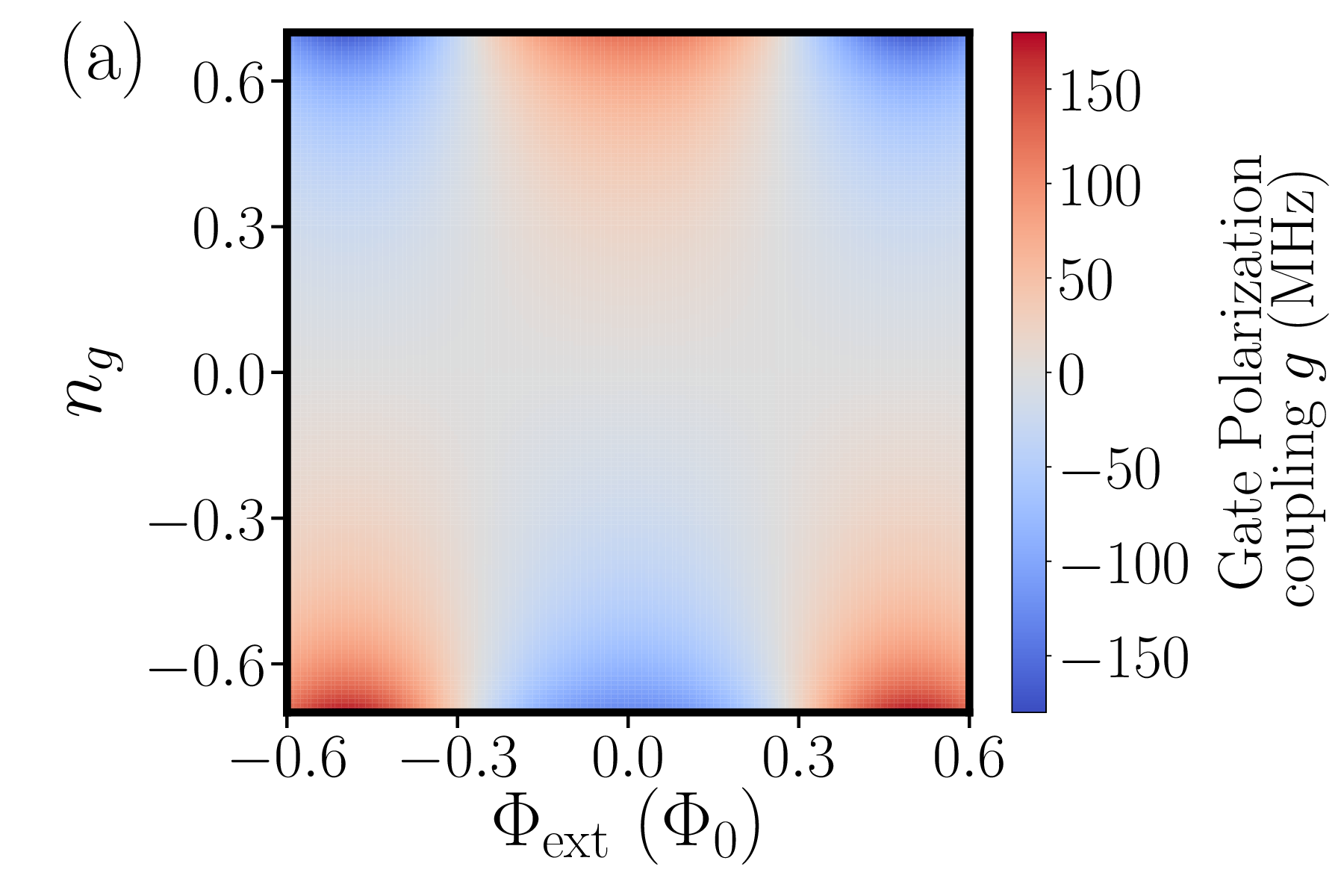}}
\hspace{0.6cm}
\subfloat{\label{g_wo_ratio}\includegraphics[width=0.4\textwidth,trim=0 0 0 0, clip]{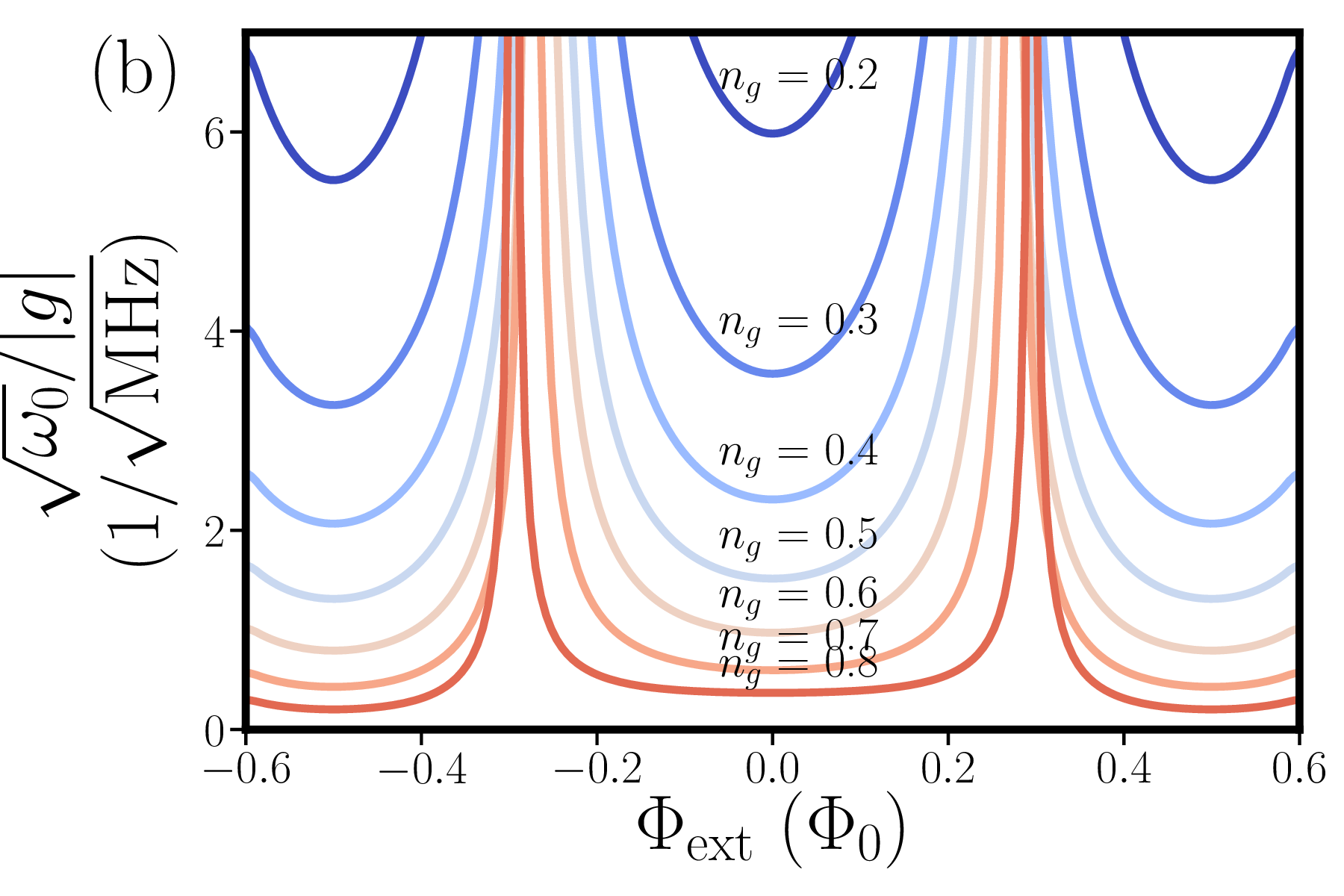}}\\
\subfloat{\label{charge_sens_color} \includegraphics[width=0.4 \textwidth,trim=0 0 0 0, clip] {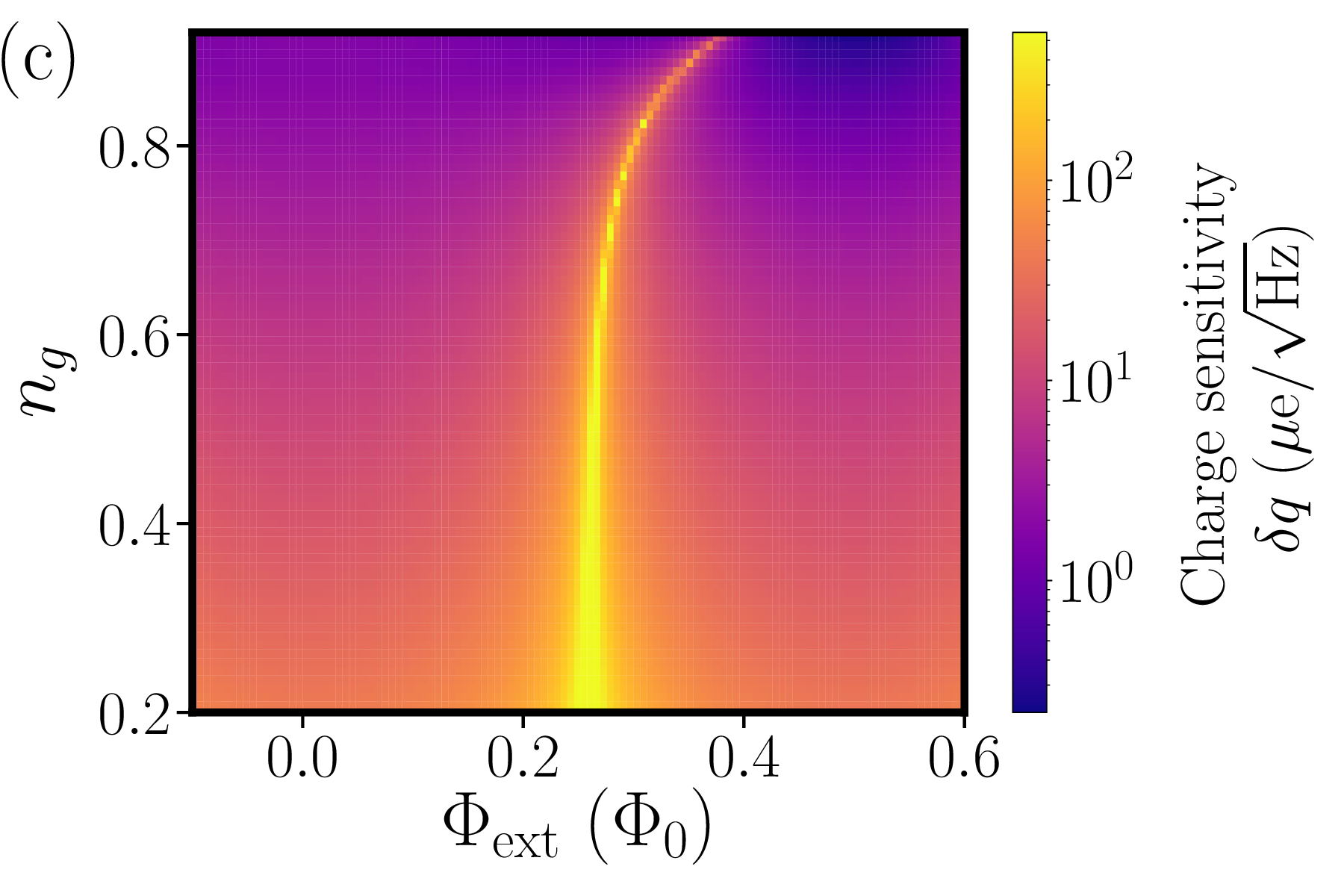}}
\hspace{0.6cm}
\subfloat{\label{wg_kappa_sat} \includegraphics[width=0.4 \textwidth,trim=0 0 0 0, clip] {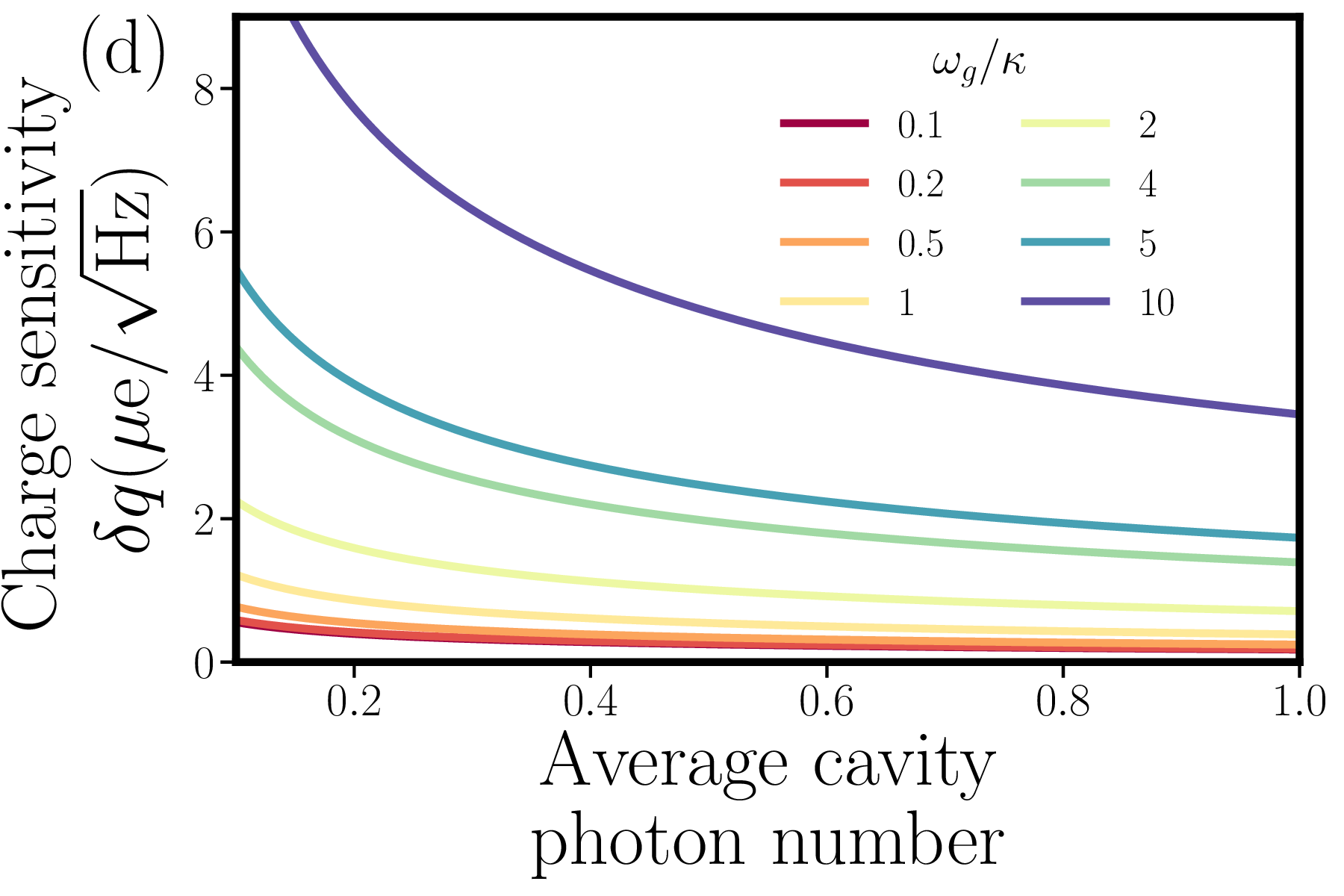}}
\caption{(a) Gate polarization coupling in MHz across the tunable bias range. The coupling becomes stronger in the direction of charge degeneracy. (b) The ratio $\sqrt{\omega_0}/|g|$ as a function of $\Phi_{\text{ext}}$ for different values of $n_g$. The fundamental charge sensitivity $\delta q$ is proportional to this ratio and the improved values are attained closer to charge degeneracy. (c) $\delta q$ for an average of one photon in the cavity, with $\omega_g/\kappa = 1$. The values reported here assume  contribution from a single side-band. (d) Comparing $\delta q$ in the bad-cavity and good-cavity limit. The bias point is chosen at $(\Phi_{\text{ext}} = 0.5 \Phi_0, n_g = 0.9)$ which gives $\delta q = 0.17 \; \mu e / \sqrt{\text{Hz}}$ for an average of one photon in the cavity in the bad-cavity limit.}\label{charge_sens_fig}
\end{figure*}

The output power at the sample stage in the presence of a gate modulated signal can be estimated using the same series of steps as for the bare cavity in Sec. \ref{bare}. In particular, we proceed to derive a modified quantum Langevin equation (\ref{ftql3eq}) and then extend the resulting input-output equation to find the analogous expression to Eq. (\ref{outpower_barecav}) that represents the output power (\ref{signalnoiseeq}). Details of this derivation are given in the Appendix \ref{output_power_appendix}, where we observe from Eqs. (\ref{aeq})-(\ref{iterateq}) that the gate modulation introduces sidebands into the cavity frequency spectrum, and is detected by measuring the output power as expressed in Eq. (\ref{poutexp}). 

Internal noise/losses  are modeled as a second, internal thermal bath denoted as $\rho_{{{\iota}}}$, modifying the total input state: $\rho_{\text{in}} = \rho_{\alpha,p} \otimes \rho_{{{\iota}}}$.
%associated with an `in' operator denoted as $a_{{\iota}}^{\text{in}}(\omega)
The thermal occupancies of the pump $n_p$ and internal bath $n_{{\iota}}$ are usually assumed to be identical, as the temperature variations at different locations in the device are neglected. However, in reality, the internal bath may have a different noise temperature due, for example, to coupling with two-level defects.\cite{martinis1993effect}

The major motivation behind the theoretical framework provided in this paper is to identify the potential applicability and fundamental limitations of the cCPT as a linear charge detector subject to the laws of quantum mechanics. This essentially implies disregarding the sources of noise that may arise from any experimental materials complexity and which are not limited in principle by quantum mechanics. To address this fundamental charge sensitivity limit, we shall therefore neglect the internal bath by setting $\kappa_{\text{int}} = 0$, and consider the response of the cCPT at absolute zero temperature for the pump/probe line, i.e., $n_p=0$. The cCPT performance under these conditions is determined by its essential coupling with the pump/probe line at the output and the measured system at the input. In the absence of a physical system at the input, the noise feeding the input of the subsequent amplifier stage thus originates from the vacuum photon shot noise of the transmission line, determining the lower bound for the charge-sensitivity. In reality, additional noise source channels can prevent achieving this fundamental charge sensitivity limit, as discussed in detail in Sec. \ref{conclusion}.

The charge sensitivity $\delta q$ ($\mathrm{e/\sqrt{Hz}}$) of an electrometer is defined as the rms charge modulation amplitude that corresponds to a signal-to-noise ratio of one (in a bandwidth of $1~{\mathrm{Hz}}$) at the amplifier input.\citep{tosi-design-2019} We can thus solve for the fundamental charge sensitivity of the cCPT from the total output power expression (\ref{signalnoiseeq}) by setting $\omega_p = \omega_0$, and looking at the output power variation about $\omega_0 \pm \omega_g$ within a bandwidth of $\Delta \omega = 2 \pi \times 1$ Hz to obtain:
\begin{equation}
    \delta q= |g|^{-1} \sqrt{\frac{\hbar\omega_0 \left(\omega_g^2+(\kappa/2)^2\right)}{4P_p^{\mathrm{in}}}} \, \mathrm{e/\sqrt{Hz}},
    \label{fundchargesenseq}
\end{equation}
where $\kappa$ now denotes the damping solely due to the coupling to pump/probe line $\kappa_{\mathrm{ext}}$, given by Eq. (\ref{cavdampeqn}) in terms of the cCPT renormalized fundamental resonance frequency (\ref{rencavfreqeq}). Equation (\ref{fundchargesenseq}) may alternatively be expressed in terms of the average photon number in the cavity $n_{\mathrm{cav}}$ as follows:
\begin{equation}
    \delta q= |g|^{-1} \sqrt{\frac{ \left(\omega_g^2+(\kappa/2)^2\right)}{\kappa \;n_{\mathrm{cav}}}} \, \mathrm{e/\sqrt{Hz}}.
\end{equation}

The sensitivity may be further improved using a homodyne detection scheme, where the combined contribution of both the sidebands lead to values lower by a factor of $\sqrt{2}$.\cite{brenning-ultrasensitive-2006}

The most charge sensitive points can be identified using the plots in Fig. \ref{charge_sens_fig}. Regardless of the input drive and signal frequency $\omega_g$, the charge sensitivity in general improves as $n_g$ approaches (but does not equal) one [Fig. \ref{charge_sens_fig}(b)]. In the case of an average of one photon in the cavity with $\omega_g/\kappa =1$, Fig. \ref{charge_sens_fig}(c) shows the  fundamental charge sensitivity behavior across the entire bias range for a single sideband. We obtain $\delta q =0.39 \, \mathrm{\mu e/\sqrt{Hz}}$ at $(\Phi_{\mathrm{ext}}, n_g) = (0.5 \, \Phi_0, 0.9)$ for the above parameter values, while working well within the adiabatic approximation limit. Moreover, the efficiency of the charge detector can be best exploited in the bad-cavity limit $\omega_g \ll \kappa$, where the value of $\delta q$ saturates to $0.17 \, \mathrm{\mu e/\sqrt{Hz}}$ for an average of one cavity photon [Fig. \ref{charge_sens_fig}(d)]. The values used in our numerical simulations are close to the experimental ones reported in Ref. \onlinecite{brock-ccpt-2020}; however an optimization of the $E_C,E_J$ values may further improve the charge sensitivity slightly.

It is worthwhile noting that the highly anharmonic, effective potential (\ref{potential_ten}) of the cCPT leads to non-negligible contributions from the quartic Kerr potential term even near an average of one cavity photon. In theory, it is possible to substantially improve the performance of the cCPT  by driving the cavity at the onset of bistability (and where the cCPT still behaves as a linear electrometer) as long as the signal is within $g \delta n_g^{(0)}/ \omega_g \ll 1$.\cite{laflamme2011quantum,tosi-design-2019}

\section{Conclusion}
\label{conclusion}

One of the key applications of the cCPT is to perform quantum measurements using phase-preserving amplification of an observable of another measurable quantum system, such as a qubit or a mechanical resonator. Of particular interest is such a tripartite coupling involving the cavity and a mechanical resonator interacting via the CPT, where the resulting, tunable CPT-induced effective optomechanical interaction may approach the single photon-single phonon ultrastrong coupling regime.\citep{rimberg-cavity-cooper-2014}

Since the device operation is limited by quantum noise, a natural extension of the present work is to investigate how close the cCPT detector approaches the standard quantum limit, with the back-action of the cCPT on the measured system taken into account. In the conventional case of large photon driving, the  coupling term in the opto-mechanical Hamiltonian can be linearized in the cavity and mechanical oscillator coordinates, and the information about the position of the mechanical resonator can be extracted using a single quadrature measurement.\citep{bowen-quantum-2016} As a result, the uncertainty in the back-action noise $S_{FF}(\omega)$ and the imprecision noise in position $S^{\text{imp}}_{xx}(\omega)$ are bounded by the inequality $S_{FF} S_{xx}^{\text{imp}} \geq \hbar/2$. In the low average cavity photon number limit, however, we must retain the original form of the opto-mechanical Hamiltonian:\citep{nunnenkamp-single-photon-2011}
\begin{eqnarray}
\label{optomech_ham_eqn}
    \mathcal{H} &=& \hbar \Delta a^{\dagger} a + \hbar \Omega \; b^{\dagger} b + \hbar G a^{\dagger} a x,
    \end{eqnarray}
    where $a$ and $b$ denote the cavity and mechanical resonator annihilation operators respectively, $x$ is the oscillator position, $\Delta = \omega_0 - \omega_p$, $\Omega$ is the mechanical oscillator frequency, and $G$ determines the opto-mechanical coupling. As a result, the radiation pressure force power spectral density is given by
    \begin{equation}
      S_{FF}(\omega) =  \left(\frac{\hbar G}{x_{zp}} \right)^2 S_{NN} (\omega),
    \end{equation}
    where $S_{NN}$ is the cavity photon number noise and $x_{zp}$ is the mechanical resonator position zero-point uncertainty. Hence we expect a quantum-limited inequality with  imprecision noise depending on the phase noise $S_{\theta \theta}(\omega)$. Investigations probing the standard quantum limits achievable in the combined cCPT-mechanical oscillator system in the presence of low average photon number drive thus requires considering ways to measure the phase operator itself. It is worthwhile noting that the typical approximation for the phase in terms of the quadratures, $\langle\hat{\theta}\rangle = \langle\hat{Y}\rangle/\langle\hat{X}\rangle$, no longer holds in this limit;\citep{clerk-introduction-2010}  further studies at a fundamental level are required to understand the behaviour of the phase operator, both theoretically and experimentally.\citep{ carruthers-phase-1968,shapiro-quantum-1991,gerhardt-ideal-1973,skagerstam-quantum-2004, barnett-hermitian-1989,gerry-introductory-2004}

Despite the cCPT's potential as an ultra-sensitive charge detector, the experimental limitations during fabrication and measurements can hinder its ability to perform at optimum sensitivity.\citep{brock2021fast} In addition to the noise contributions at the sample stage, the measurement precision is also limited by the noise added at the subsequent amplifier stages (where the minimum noise added by a quantum-limited phase insensitive amplifier is $\hbar \omega_0/2$). As we mentioned, other transport mechanisms such as quasiparticle poisoning may dominate the resonance characteristics when we operate closer to charge degeneracy. The internal damping of the cavity further limits the charge sensitivity, modifying the fundamental, quantum limited expression (\ref{fundchargesenseq}) as follows:
\begin{equation}
    \delta q=|g|^{-1} \frac{\kappa_{\text{tot}}}{\kappa_{\text{ext}}}  \sqrt{\frac{\hbar\omega_0 \left(\omega_g^2+(\kappa_{\text{tot}}/2)^2\right)}{4P_p^{\mathrm{in}}}} \, \mathrm{e/\sqrt{Hz}}.
    \label{chargesenseq}
\end{equation}

Most importantly, the increased sensitivity to minute variations in the gate voltage also makes the cCPT prone to gate charge fluctuations that are potentially due to the two-level fluctuators arising within the thin oxide layers of the device.\citep{paladino-mathbsf1mathbsfitf-2014} This incoherent coupling results in resonant frequency fluctuations during real-time measurements that are typically manifested as $1/f$ noise, which make it challenging to precisely set the pump tone on resonance as we have assumed. While there exist several detection techniques for the measurement of such low-frequency noise,\citep{neill-fluctuations-2013,lindstrom-pound-locking-2011,brock-frequency-2020} methods to suppress these fluctuations in real-time are at present under development \citep{poundlocking_print}; the suppression of such noise could potentially lead to major breakthroughs in several areas of research, ranging from charge detection to applications in qubit metrology.\citep{muller-towards-2019,paladino-mathbsf1mathbsfitf-2014,yang-achieving-2019}

In this paper, we have presented a first principles, theoretical model of a quantum-limited linear electrometer. 
%With the addition of a gate voltage coupled nanomechanical resonator, the cCPT electrometer is in principle capable of measuring macroscopic opto-mechanical states in the single photon-phonon coupling regime \cite{rimberg-cavity-cooper-2014}. 
The model uses adiabatic elimination of the CPT dynamics, such that the cCPT passively mediates the interactions between the microwave cavity and the measured system (e.g., mechanical resonator) via linear charge sensing. For parameters similar to those of the experimental device described in Ref. \onlinecite{brock-ccpt-2020}, we predict the fundamental, quantum noise limited charge sensitivity of the cCPT linear electrometer to be $ 0.12 \, \mu$e/$\sqrt{\mathrm{Hz}}$ under a homodyne detection scheme. This sensitivity corresponds to the pumped cavity having an average of one photon, with the cCPT operated in the gate tunable range $0 \leq n_g \leq 0.9$, where the adiabatic approximation is valid and the effects of quasiparticle poisoning may be reduced in an experimental device.

\begin{acknowledgments}
We thank Bhargava Thyagarajan, 
William Braasch, Josh Mutus and Juliang Li for very helpful discussions. This work was supported by the NSF under Grants No. DMR-1807785 (S. K., B. L. B., and A. R) and  DMR-1507383 (M. B.), and by an unrestricted gift from Google (S. K.).
\end{acknowledgments}

\appendix
\section{Lumped element circuit analysis}
\label{circuit}
%The cavity-embedded Cooper pair transistor (cCPT) consists of a shorted quarter-wave ($\lambda/4$) resonator in a co-planar wave guide geometry, and a Cooper pair transistor (CPT) at the voltage anti-node. (Fig \ref{ccpt_scheme}). 
As discussed in detail in Sec \ref{cCPT}, the CPT leads to a tunable quantum inductance, and can be modeled as a nonlinear inductor in parallel with the bare cavity,  which in turn leads to a tunable resonance. Here we discuss the CPT characteristics employing a lumped element circuit analysis, which can be used to provide check-points for the first principles, operator scattering analysis, under appropriate limits. %In this section we discuss the CPT characteristics employing a lumped element circuit analysis, which we later use to provide check-points for the first principles field operator analysis, under appropriate limits.

%To begin, 
The frequency response near resonance for a bare cavity (Fig \ref{barec_scheme})  depends on the input impedance $Z_{\text{in}}^{\lambda/4}$ given by
\begin{equation}
\label{input_impedance_eqn_quarterwave}
    Z_{\text{in}}^{\lambda/4} (\delta x) \approx\frac{4 Z_0 Q_{\text{int}} / \pi}{1 + 2iQ_{\text{int}} \delta x},
\end{equation}
where $Z_0$ is the characteristic impedance, $Q_{\text{int}}$ is the quality factor representing internal losses and $\delta x = (\omega - \omega_{\lambda/4})/ \omega_{\lambda/4}$.  Here, $\omega_{\lambda/4}=(2n+1)\pi v_c/2l,\, n=0,1,2\dots$ are the resonant frequencies obtained by applying the shorted quarter-wave condition of the bare cavity, with $v_c$ the phase velocity. For an attenuation constant $\alpha$, this input impedance is equivalent to that of a parallel RLC circuit with resistance $R_{\text{cav}}=Z_0/\alpha l$, cavity mode capacitance $C_{\text{cav}}=\left(4 Z_0 \omega_{\lambda/4} / \pi \right)^{-1}$, and cavity mode inductance $L_{\text{cav}}=\left(\omega_{\lambda/4}^2 C_{\text{cav}}\right)^{-1}$.\citep{david-m-pozar-microwave-2012}

The extraction of scattering parameters is achieved using reflection measurements by means of a pump-probe transmission line weakly coupled to the cavity via the capacitance $C_{pc} \ll C_{\text{cav}}$. This coupling capacitance leads to an added impedance and shifted resonance $\omega_n$ obtained through the condition
\begin{equation}
    \operatorname{\mathbb{I}m} \left(Z_r\vert _{\omega = \omega_n}\right) = \operatorname{\mathbb{I}m} \left(-\frac{i}{\omega_n C_{pc}} + Z_{\text{in}}^{\lambda/4}(\omega_n) \right) = 0,
    \label{impedance_circuit}
\end{equation}
which gives
\begin{equation}
    \label{bare_cavity_w0_circuit}
    \omega_n \approx \omega_{\lambda/4} \left( 1 - \frac{C_{pc}}{2 C_{\text{cav}}} \right),\, n = 0, 1, 2\dots,
\end{equation}
 where we neglect the second possible solution owing to its high resulting impedance.

The equivalent lumped element model of the combined system thus modifies to a series RLC circuit in this configuration of weak coupling. The input impedance near resonance is\citep{megrant-planar-2012,mazin-microwave-2005}
\begin{equation}
    \label{eqn}
    Z_r(\omega) \approx Z_0 \frac{Q_{\text{ext}}}{Q_{\text{int}}} \left( 1+2 i Q_{\text{int}} \frac{\Delta \omega}{\omega_{n}}\right),
\end{equation}
with parameters $R_{\text{cav}}=Z_0 Q_{\text{ext}}/Q_{\text{int}}$, $L'_{\text{cav}}=Z_0 Q_{\text{ext}} /\omega_{n}$ and $C'_{\text{cav}}= \left( \omega_{n}^2 L'_{\text{cav}} \right)^{-1}$, where the external probe coupling quality factor is obtained using
\begin{equation}
    Q_{\text{ext}}(\omega)=\omega \, \frac{\text{Energy stored}}{\text{Power loss}}
    = \frac{\pi}{4 (\omega C_{pc} Z_0)^2}\frac{\omega}{\omega_{\lambda/4}}
    \label{bare_cavity_qc}.
\end{equation}

For the parameters of our experimental device, the relative variation in the external quality factor is small near resonance:  $\Delta Q_{\text{ext}}(\omega)/ Q_{\text{ext}}(\omega_0) \ll 1$.  We therefore approximate the external damping rate to be constant over the frequency region of interest. This allows us to work under the Markovian approximation, which considerably simplifies the calculations.
 
Equation (\ref{bare_cavity_w0_circuit}) for the cavity resonant frequency  can be reexpressed in terms of a renormalized total capacitance $C_{\text{cav}} \to C_{\text{cav}} + C_{pc}$. Hence, the addition of the CPT shifts the resonance via an effective capacitance $C_{\text{cCPT}} = C_{\text{cav}} + C_{pc} + C_{\text{CPT}}$, and an effective inductance
 $L_{\text{cCPT}}^{-1} = {L}_{\text{cav}}^{ -1} \,+ \, L_{\text{CPT}}^{-1}$.
 Consequently, under the conditions $C_{\text{CPT}}/C_{\text{cav}},\,  L_{\text{cav}}/L_{\text{CPT}} \ll 1$, the cCPT resonant frequency is renormalized to
 \begin{equation}
 \label{ccpt_res_circuit}
     \omega_{\text{cCPT}} \approx \omega_{\lambda/4} \left(1 +\frac{ L_{\text{cav}}}{2L_{\text{CPT}}} - \frac{C_{pc} + C_{\text{CPT}}}{2C_{\text{cav}}} \right).
 \end{equation}
 
%The device is also designed such that $L_{\text{CPT}}$ has a wide tunability range via two parameters (see Sec. \ref{cCPT}): the voltage $V_g$ gating the CPT island  and  the external magnetic flux $\Phi_{\text{ext}}$ threading the closed loop formed by the CPT, shorted center conductor, and ground plane (Fig. \ref{ccpt_scheme}). 

\section{Quantum Langevin equation for the bare cavity}
\label{bare_cavity_appendix}
For completeness, here we verify that the  standard quantum Langevin equation in the Fourier domain can be obtained using the results presented in Sec. \ref{fullanalysis}. Furthermore, we can extract the familiar closed-system cavity mode Hamiltonian, and the zero-point fluctuations of the cavity phase modes.

Simplifying Eq. (\ref{cavityasolneq}) by approximation using (\ref{bare_cavity_w0_field})  and restricting to a narrow bandwidth $\Delta\omega\ll\omega_n$, we obtain to first order in the capacitance ratio $\xi \equiv C_{pc}/({\mathcal{C}}_c l)\ll 1$: 
\begin{equation}
\left(\omega-\omega_n +i\frac{\kappa_{\text{ext}}}{2}\right) a_n(\omega)=\sqrt{\kappa_{\text{ext}}} \; a_p^{\mathrm{in}}(\omega).
\label{ftql2eq}
\end{equation}
This expression is the standard, Fourier transformed quantum Langevin equation, where the $n$th cavity mode photon annihilation operator is defined as
\begin{equation}
    a_n(\omega) \equiv \sqrt{\frac{2l}{v_c}}e^{i\omega t_0} a_c^\to (\omega,t_0),
    \label{arescaleq}
\end{equation}
for $\omega$ in the vicinity of a given mode frequency $\omega_n$ [Eq. \ref{bare_cavity_w0_field})].
This rescaling ensures that $a_n (t)=\frac{1}{\sqrt{2\pi}}\int_{-\infty}^{+\infty} d\omega e^{-i\omega t} a_n(\omega)$ satisfies the usual, discrete mode canonical commutation relation $[a_n(t),a^{\dag}_n(t)]=1$.
%Furthermore, under the Markovian approximation, the pump/probe damping rate $\kappa_{\text{ext}}$ matches (\ref{bare_cavity_qc}) near $\omega_n$, with $Q_{\text{ext}} = \omega_n / \kappa_{\text{ext}}$, and is given by
%\begin{equation}
    %\kappa_{\text{ext}} = 2Z_p\frac{C_{pc}^2}{{\mathcal{C}}_c l}\omega_n^2.
    %\label{cavdampeqn}
%\end{equation}

The Hamiltonian of the closed system consisting of a shorted quarter-wave resonator with a coupling capacitance thus comprises discrete harmonic oscillator modes:
\begin{eqnarray}
    H_{\text{cav}} &=& \sum_{n=0}^{\infty} \hbar \omega_n \left(a_n^{\dagger} a_n+\frac{1}{2}\right) \cr &=& \sum_{n=0}^{\infty} \left[\left(\frac{2\pi}{\Phi_0}\right)^2 \frac{p_{n}^2}{2 C_{n}} +\left(\frac{\Phi_0}{2\pi}\right)^2 \sum_{n=0}^{\infty} \frac{\phi_n^2}{2 L_{n}}\right],
    \label{barecavity_ham}
\end{eqnarray}
where we use the notation $a_n$ for the mode `$n$' cavity operator. The second line represents the Hamiltonian for the independent lumped element $LC$ oscillators expressed in terms of the generalized mode phase coordinates and  conjugate momenta respectively: $\phi_{n}=\phi_{zp,n}\left(a_{n}+a^{\dag}_{n}\right)$ and $p_n=-i\left(\Phi_0/2\pi\right)^2 \omega_n\phi_{zp,n} \left(a_n-a_n^{\dag}\right)$. The lumped element parameters are given by the mode capacitance $C_{n} = \mathcal{C}_c l/2 + C_{pc}$ and the mode inductance $L_{n} = 8 \mathcal{L}_c l / (2n+1)^2 \pi^2$, and the mode zero-point uncertainty can be written as
\begin{equation}
    \phi_{zp,n} = \left( \frac{2 \pi} {\Phi_0} \right) \sqrt{\frac{\hbar} {2 C_n \omega_n}} = 2 \sqrt{\frac{\mathcal{Z}_n}{R_K}},
    \label{zpteq}
\end{equation}
with $\mathcal{Z}_n = \pi \sqrt{L_n/C_n}$ the cavity mode impedance and $R_K = h/e^2$ the von Klitzing constant.

\section{Effective cavity quantum dynamics}
\label{ccpt_appendix}
In this appendix, we provide details of the derivation for the tunable resonance of the cCPT, following the same operator scattering method steps as utilized for the bare cavity case (Sec. \ref{fullanalysis}), but now with the boundary condition (\ref{phi4ppeq}) replacing the simpler, bare cavity boundary condition (\ref{cpjunctioneq}).  

For a sinusoidal gate modulation frequency $\omega_g \ll \omega_0$ and amplitude $\delta n_g^{(0)} \ll 1$, the term in the RHS of Eq. (\ref{phi4ppeq}) can be neglected.
Under these assumptions, we proceed by Taylor expanding the term $\partial E_{\mathrm{CPT}}^{(0)}/\partial \phi_c$ in Eq. (\ref{phi4ppeq}) to obtain:
\begin{eqnarray}
&&\phi_c'(0,t)-\frac{C_{\text{CPT}}}{{\mathcal{C}}_c }\phi_c''(0,t)-\frac{\mathcal{L}_c}{\mathcal{L}_p}\phi_p'(0,t)\cr 
&&-\left(\frac{2\pi}{\Phi_0}\right)^2 {\mathcal{L}}_c \sum_{n=1}^{\infty} \; \sum_{k=0}^n \frac{1}{n!} {n\choose k} \phi_c(0,t)^{k} \cr
&&\times
\left.\frac{\partial^{n+1} E_{\mathrm{CPT}}^{(0)}}{\partial \phi_c^{k+1} \partial n_g^{n-k}}\right|_{n_g=n_g^{(0)},\phi_c = 0} \delta n_g^{n-k} = 0,
\label{phi5ppeq}
\end{eqnarray}
where ${n \choose k}$ is the binomial coefficient, $n_g(t) = n_g^{(0)} + \delta n_g(t)$ and the gate modulation $\delta n_g(t) = \delta n_g^{(0)} \cos\left(\omega_g t\right)-N^{\mathrm{in}}_p(t)$.

Utilizing the operator scattering solutions in Eq. (\ref{phasec2solneq}) for the cavity phase field  and in Eq. (\ref{pumpphasepsoln3eq}) for the pump phase field, we arrive at the following modified pump-cavity coupled equation in frequency space:
\begin{eqnarray}
&&\left\{\cos\left(\omega l/v_c\right) -\omega  Z_c\left[\frac{ C_{pc}}{1+\left(\omega Z_p C_{pc}\right)^2}+C_{\text{CPT}}\right.\right.\cr
&&\left.\left.-\omega^{-2} \left(\frac{2\pi}{\Phi_0}\right)^2 \left. \frac{\partial^2 E_{\mathrm{CPT}}^{(0)}}{\partial \phi_c^2}\right|_{\phi_c = 0} \right] \sin\left(\omega l/v_c\right) \right\}a_c^{\to}(\omega,t_0)
\cr
&&-i\frac{\left(\omega \sqrt{Z_c Z_p} C_{pc}\right)^2 }{1+\left(\omega Z_p C_{pc}\right)^2} \sin\left(\omega l/v_c\right) a_c^{\to}(\omega,t_0) \cr 
&&= -ie^{-i\omega \left(t_0 +l/v_c\right)} \frac{\omega\sqrt{Z_p Z_c}  C_{pc} }{1-i\omega Z_p C_{pc}} a^{\mathrm{in}}_p (\omega)\cr
&&-Z_c \left(\frac{2\pi}{\Phi_0}\right)^2 \left.\sum_{n=2}^{\infty} \frac{1}{(n-1)!} \frac{\partial^{(n+1)} E_{\mathrm{CPT}}^{(0)}}{\partial \phi_c^2 \partial n_g^{(n-1)}}\right|_{n_g=n_g^{(0)},\phi_c=0} \cr 
&&\times\frac{1}{2\pi}\int_{-\infty}^{+\infty}dt \int_0^{\infty}\frac{d\omega'}{\sqrt{\omega\omega'}}e^{i\left(\omega-\omega'\right)\left(t-t_0-l/v_c\right)}\cr&&\times\delta n_g(t)^{n-1} \sin\left(\omega' l/v_c\right) a_c^{\to} (\omega',t_0) + {\mathcal{O}}(\phi_c^2),
\label{cavityasoln2eq}
\end{eqnarray}
where we have limited the expansion to first order in $\phi_c$, leaving out  anharmonic terms. As for the bare cavity case [Eq. (\ref{pumpphasesolneq})], the renormalized frequency due to the CPT and transmission line coupling can be obtained by equating the terms in curly brackets to zero. The third line corresponds to the cavity damping rate due to  coupling to the transmission line, the fourth line describes the transmission line noise, and the remaining term gives the gate voltage and noise modulations of the cavity frequency.

Defining the dimensionless frequency as $\tilde{\omega} \equiv \omega l/v_c$ and the small dimensionless CPT-transmission line coupling parameter $\xi=v_c C_{pc} Z_c/l=C_{pc}/{\mathcal{C}}_c l\ll 1$, we can express the term in curly brackets as
\begin{equation}
    \cos\tilde{\omega} -\left[\frac{\tilde{\omega}\xi}{1+\left(\tilde{\omega}\xi\right)^2}+\tilde{\omega}\frac{C_{\mathrm{CPT}}}{{\mathcal{C}}_c l}- \tilde{\omega}^{-1}\frac{{\mathcal{L}}_c l}{L_{\mathrm{CPT}}}  \right]\sin\tilde{\omega},
    \label{resfreqeq}
\end{equation}
with the CPT inductance $L_{\text{CPT}}$  defined as
\begin{eqnarray}
    %L_{\mathrm{CPT}}^{-1}= \left(\frac{2\pi}{\Phi_0}\right)^2\left.\frac{\partial^2 E_{\mathrm{CPT}}^{(0)}}{\partial \phi^2}\right|_{(n_g,2 \pi \Phi_{\text{ext}}/\Phi_0)}.
    L_{\mathrm{CPT}}^{-1}= \left(\frac{2\pi}{\Phi_0}\right)^2 \left.\frac{\partial^2 E_{\mathrm{CPT}}^{(0)}}{\partial \phi_{c}^2} \right|_{\phi_c=0} = \frac{\partial^2 E_{\mathrm{CPT}}^{(0)}}{\partial \Phi_{\text{ext}}^2},
    \label{CPTLeq}
\end{eqnarray}
utilizing Eq. (\ref{displaced_cavity_phase_eqn}).
 Setting expression (\ref{resfreqeq}) to zero, and in the limit where the CPT weakly perturbs the cavity fundamental resonance, i.e., ${C_{\mathrm{CPT}}}/{{\mathcal{C}}_c l}$ and ${{\mathcal{L}}_c l}/{L_{\mathrm{CPT}}} \ll 1$ (Fig \ref{lcptratio}), we obtain the following expression for the tunable resonance:
\begin{equation}
    \omega_0(n_g,\Phi_{\text{ext}}) \approx \frac{\pi v_c}{2l}\left[1-\frac{C_{pc}+C_{\mathrm{CPT}}}{{\mathcal{C}}_c l}+\left(\frac{2}{\pi}\right)^2\frac{{\mathcal{L}}_c l}{L_{\mathrm{CPT}}}\right].
    \label{rencavfreqappeq}
\end{equation}

\section{Output power for a gate-modulated signal}
\label{output_power_appendix}
This appendix details how a sinusoidal modulated signal at the CPT's gate introduces side-bands into the frequency spectrum of the output power, measured via the pump/probe transmission line. 

Limiting the relevant frequency space to the region of the fundamental cavity mode frequency: $|\omega-\omega_0|\ll\omega_0$, we obtain from Eq. (\ref{cavityasoln2eq}) the following, approximate modified quantum Langevin equation to first order in $\xi=C_{pc}/({\mathcal{C}}_c l)\ll 1$:
\begin{eqnarray}
    &&\left(\omega-\omega_0 +i\frac{\kappa_{\text{ext}}}{2}\right) a_0(\omega)=\sqrt{\kappa_{\text{ext}}} a_p^{\mathrm{in}}(\omega)\cr && + g \int_0^{\infty} d\omega'F(\omega, \omega'),
    \label{ftql3eq}
\end{eqnarray}
where $a_0$ is given by Eq. (\ref{arescaleq}) for $\omega$ in the vicinity of the cCPT renormalized, fundamental mode frequency $\omega_0$ [Eq. (\ref{rencavfreqeq})] and $\kappa_{\text{ext}}$ is given by Eq. (\ref{cavdampeqn}) similarly in terms of the cCPT renormalized fundamental resonance frequency. Note that the
 gate modulation introduces higher order corrections to $a_0(\omega)$ via the term
\begin{eqnarray}
    &&F(\omega,\omega') = \frac{ 1}{\sqrt{2\pi}}\frac{ \omega_0}{\sqrt{\omega \omega'}}e^{-i\left(\omega-\omega'\right)l/v_c}
    \cr&&\times\biggl\{\sqrt{\frac{\pi}{2}}\delta n_g^{(0)} \left[\delta\left(\omega-\omega'+\omega_g\right)+\delta\left(\omega-\omega'-\omega_g\right)\right]
    \cr&&-N^{\mathrm{in}}_p(\omega-\omega')\biggr\}\sin\left(\omega' l/v_c\right) a_0(\omega').
    \label{feq}
\end{eqnarray}

We may further simplify Eq. (\ref{ftql3eq}) by neglecting $N^{\mathrm{in}}_p(\omega-\omega')$ in Eq. (\ref{feq}) owing to the smallness of its noise contribution, and noting also that the $\omega,\, \omega'$ dependent terms multiplying $a_0(\omega')$ in Eq. (\ref{feq})  can be approximately evaluated at $\omega_0$ since we assume $\omega_g,\kappa_{\mathrm{ext}},\Delta\omega\ll \omega_0$, where $\Delta\omega$ is the measured output power bandwidth centered at the pump frequency $\omega_p$.  Introducing internal effective cavity losses using a phenomenological constant damping rate  $\kappa_\text{int}$, channeled via an additional non-measurable input port $a_{\iota}^{\text{in}}(\omega)$, we obtain:
\begin{eqnarray}
    &&\left(\omega-\omega_0 +i\frac{\kappa_{\text{tot}}}{2}\right) a_0(\omega)=\sqrt{\kappa_{\text{ext}}} a_p^{\mathrm{in}}(\omega)+\sqrt{\kappa_{\text{int}}} a_{{\iota}}^{\mathrm{in}}(\omega)\cr
&+& g\int_0^{\infty}{{d\omega'}} A(\omega - \omega') a_0(\omega'), 
\label{ftql4eq}
\end{eqnarray}
where 
\begin{equation}
    A(\omega) = \frac{1}{2}\delta n_g^{(0)} \left[\delta\left(\omega+\omega_g\right)+\delta\left(\omega-\omega_g\right)\right].
    \label{aeq}
\end{equation}

Solving Eq. (\ref{ftql4eq}) perturbatively in the limit of small $g$, we have
\begin{equation}
    a_0({\omega}) =  \sum_{n=0} g^n \; \mathcal{I}_{n} (\omega),
    \label{acsolneq}
\end{equation}
where the zeroth order term in (\ref{acsolneq}) is 
\begin{equation}
\mathcal{I}_0(\omega) = \frac{\sqrt{\kappa_{\text{ext}}} a_p^{\mathrm{in}}(\omega)+\sqrt{\kappa_{\text{int}}} a_{{\iota}}^{\mathrm{in}}(\omega)}{\left(\omega-\omega_0 +i\frac{\kappa_{\text{tot}}}{2}\right)},
\end{equation}
and the iterative solution relation for $\mathcal{I}_n(\omega)$ is given by
\begin{eqnarray}
    \mathcal{I}_n(\omega) = \int_0^{\infty} d\omega' \; \frac{\mathcal{I}_{n-1} (\omega') A(\omega - \omega')}{\omega - \omega_0 + i \frac{\kappa_{\text{tot}}}{2}}.
    \label{iterateq}
\end{eqnarray}

Considering the time-domain expression for $a_0(t)$ using Eq. (\ref{ftql4eq}), we obtain the following necessary condition for linear charge detection: $g \delta n_g^{(0)}/ \omega_g \ll 1$. In this linear detection regime, the output power reaching the first-stage amplifier is given by
\begin{eqnarray}
&&P^{\mathrm{out}}(\omega_0,\Delta\omega)=P_p^{\mathrm{in}}\int_{\omega_0-\Delta\omega/2}^{\omega_0+\Delta\omega/2}d\omega \cr && \Biggl\{\frac{\delta\omega^2+\left(\frac{\kappa_{\text{ext}}-\kappa_{\text{int}}}{2}\right)^2}{\delta\omega^2+\left(\frac{\kappa_{\mathrm{tot}}}{2}\right)^2}\delta(\omega-\omega_p) +\frac{\left(\kappa_{\text{ext}} g\delta n_g^{(0)}/{2}\right)^2}{\delta\omega^2+\left(\frac{\kappa_{\mathrm{tot}}}{2}\right)^2}\cr
&&\times\Biggl[\frac{1}{\left(\delta\omega+\omega_g\right)^2+\left(\frac{\kappa_{\mathrm{tot}}}{2}\right)^2} \delta(\omega+\omega_g-\omega_p) \cr && +\frac{1}{\left(\delta\omega-\omega_g\right)^2+\left(\frac{\kappa_{\mathrm{tot}}}{2}\right)^2} \delta(\omega-\omega_g-\omega_p)\Biggr]\Biggr\}\cr&&
%+\frac{\hbar\omega_0}{2\pi}\int_{\omega_0-\Delta\omega/2}^{\omega_0+\Delta\omega/2}d\omega \frac{\kappa_{\text{ext}}\left(\frac{g\delta n_g^{(0)}}{2}\right)^2}{\left(\omega-\omega_c\right)^2+\left(\frac{\kappa_{\mathrm{tot}}}{2}\right)^2}\cr &&
%\left\{\frac{\kappa_{\text{ext}}\left[n_p(\omega+\omega_g)+\frac{1}{2}\right]+\kappa_{\text{int}}\left[n_e(\omega+\omega_g)+\frac{1}{2}\right]}{\left(\omega-\omega_c+\omega_g\right)^2+\left(\frac{\kappa_{\mathrm{tot}}}{2}\right)^2}\right.\cr
%&&\left.+\frac{\kappa_{\text{ext}}\left[n_p(\omega-\omega_g)+\frac{1}{2}\right]+\kappa_{\text{int}}\left[n_e(\omega-\omega_g)+\frac{1}{2}\right]}{\left(\omega-\omega_c-\omega_g\right)^2+\left(\frac{\kappa_{\mathrm{tot}}}{2}\right)^2}\right\}\cr&&
+\frac{\hbar\omega_0}{2\pi}\int_{\omega_0-\Delta\omega/2}^{\omega_0+\Delta\omega/2}d\omega\left[n_p(\omega)+\frac{1}{2}\right.\cr
&&+\left.\frac{\kappa_{\text{ext}} \kappa_{\text{int}}\left(n_{\iota}(\omega)-n_p(\omega)\right)}{(\omega-\omega_p)^2+\left(\frac{\kappa_{\mathrm{tot}}}{2}\right)^2}\right].
\label{signalnoiseeq}
\end{eqnarray}
 Since $g \delta n_g^{(0)}/ \omega_g \ll 1$, we neglect the noise floor contribution of $g^2$ order. We also neglect the order $g^2$ signal contribution at $\omega=\omega_p$, which is dominated by the reflected pump tone; the actual signal is obtained from either (or both) of the sidebands at $\omega_p\pm\omega_g$.
 
\bibliography{ref}
\end{document}